\documentclass[usenatbib]{mn2e}
\usepackage{epsfig}
\usepackage{amsmath}
\usepackage{amssymb}
\usepackage{graphicx,graphics,subfigure}
\def\gtrsim{\mathrel{\raise0.35ex\hbox{$\scriptstyle >$}\kern-0.6em
\lower0.40ex\hbox{{$\scriptstyle \sim$}}}}
\def\lesssim{\mathrel{\raise0.35ex\hbox{$\scriptstyle <$}\kern-0.6em
\lower0.40ex\hbox{{$\scriptstyle \sim$}}}}

\def\h50{h_{50}}

\title[Close companions to Brightest Cluster Galaxies]{Close companions to Brightest Cluster Galaxies: Support for minor mergers and downsizing}

\author [Edwards et al.]{Louise~O.V.~Edwards$^{1,}$$^2$, David~R. Patton$^2$\\
$^1$Department of Physics, Mount Allison University,
 Sackville, NB, Canada, E4L 1E6; ledwards@mta.ca\\
$^2$Department of Physics and Astronomy, Trent University, Peterborough, ON, Canada, G1K 7P4\\
}

\date{\today}

\begin{document}
\maketitle

\begin{abstract}

We identify close companions of Brightest Cluster Galaxies (BCGs) for the purpose of quantifying the rate at which these galaxies grow via mergers. By exploiting deep photometric data from the CFHTLS, we probe the number of companions per BCG (N$_{c}$) with luminosity ratios down to those corresponding to potential minor mergers of 20:1.  We also measure the average luminosity in companions per galaxy (L$_{c}$). We find that N$_{c}$ and L$_{c}$ rise steeply with luminosity ratio for both the BCGs, and a control sample of other bright, red, cluster galaxies. The trend for BCGs rises more steeply, resulting in a larger number of close companions. For companions within 50$\,$kpc of a BCG, N$_{c}$= 1.38$\pm$0.14 and L$_{c}$=2.14$\pm$0.31$\times 10^{10}$L$_{\odot}$ and for companions within 50$\,$kpc of a luminosity matched control sample of non-BCGs, N$_{c}$=0.87$\pm$0.08 and L$_{c}$=1.48$\pm$0.20$\times 10^{10}$L$_{\odot}$. This suggests that the BCGs are likely to undergo more mergers compared to otherwise comparable luminous galaxies. Additionally, compared to a local sample of luminous red galaxies, the more distant sample presented in this study (with redshifts between 0.15-0.39,) shows a higher N$_{c}$, suggesting the younger and smaller BCGs are still undergoing hierarchical formation. Using the Millennium Simulations we model and estimate the level of contamination due to unrelated cluster galaxies. The contamination by interloping galaxies is 50\% within projected separations of 50$\,$kpc, but within 30$\,$kpc, 60\% of identified companions are real physical companions. We conclude that the luminosity of bound merger candidates down to luminosity ratios of 20:1 could be adding as much as 10\% to the mass of a typical BCG over 0.5$\,$Gyr at redshifts of z$
\sim$0.3.

\end{abstract}


\section{Introduction}

Deducing the formation and evolution of galaxies is an active area in both observational and theoretical work in astronomy. Hierarchical formation of galaxies through mergers is certainly among the main processes in a galaxy's evolutionary history. In modern studies, close galaxy pairs are used as a diagnostic of merger activity, as they should be precursors to mergers \citep[e.g.,][]{kit08}.  

Galaxies with close companions have been found to have higher asymmetry \citep{pat00,her05,pat05,dep07,mci08}, enhanced star formation \citep{lam03,alo04,nik04,bar07,ell08,li08,ell10,woo10}, bluer central colours \citep{pat11}, lower metallicities \citep{kew06,ell08} and higher AGN fractions \citep{ell11} than galaxies without close companions. This indicates that mergers can significantly affect the morphological, star forming and chemical properties of the galaxies involved in these interactions.

The frequency of close galaxy pairs can be used to determine the merger fraction, and this has been calculated from numerous studies using large galaxy surveys at various redshifts \citep[e.g.,][]{pat02,kar07,hsieh08,lin08,pat08,raw08,der09,los10,lot11}. The merger rate depends on galaxy mass, the luminosity ratio between merging galaxies and the surrounding environment \citep{kho01,ber06,cox08,guo08,pat08}. No one environment is host to merging galaxies, as galaxy groups \citep{got05,bro06,nol07}, cluster infall regions \citep{van99,tra05,mos06}, and field galaxies \citep{bar07} have all been shown to undergo mergers. \citet{ell10} find that lower density environments contain a higher fraction of pairs, but high density environments such as clusters still contain a significant population of pairs \citep{mas06,pat11}. What is more, major mergers in luminous cluster galaxies continue to occur, even at low redshift \citep{mci08,bro11}.

The period of time between z=0-1 has seen a growth of the red sequence \citep{wak06,fab07,bro07,ruh09} which is thought to be spurred by major merging events \citep{sch07,wil09}, at least in L$_{*}$ galaxies. However, the importance of minor mergers on the build up of the red sequence has recently gained attention. \citet{lot11} find that the rate of minor mergers (10:1) is three times that for major mergers (4:1). \citet{sha10} have studied the morphology and kinematics of a large sample of elliptical type galaxies and conclude that recent star forming events are likely to be the product of minor mergers that are building up the galaxy's bulge. Similarly, \citet{kav10} find too many morphologically disturbed and star forming ellipticals to be accounted for by major mergers alone. Since minor mergers are predicted to be a more common phenomena \citep{guo08}, it follows that minor mergers largely influence the build up of elliptical galaxies at late epochs. 

In the hierarchical view of galaxy formation,  the most massive galaxies would be those most recently formed. At first, this seems at odds with the idea of downsizing, where the most massive galaxies formed first. Understanding the formation and evolution of these most massive galaxies, therefore, is of particular importance. The most massive galaxies are Brightest Cluster Galaxies (BCGs), where the bulk of the stellar mass is from extremely red and very old stars. BCGs live in a particularly complex environment at the gravitational potential centre of clusters, where dense X-ray emitting intracluster gas is often found. The galaxies in these massive X-ray clusters may be old simply because they were the first density peaks of the early universe to collapse \citep{tre05}. However, their formation and evolution might not be straightforward.  Particularly intriguing are indications of recent star formation, such as blue colours and optical line emission which have been observed in BCGs that lie at the X-ray peak of cool-core clusters \citep{cra99,mcn04,edw07b,raf08,bil08}. Thus, cluster-scale properties likely factor into the presence of emission lines seen in BCGs, but other triggers may exist. Outside of cool-core clusters and BCGs, 10\% of luminous red galaxies show optical emission lines at their centres, regardless of whether the local galaxy density describes a massive cluster, or poor group \citep{edw07b}. Could this conspicuous line emission in BCGs and other luminous red galaxies be the result of recent mergers? The cool gas associated with cool-core BCGs may be influenced by interactions. \citet{don07} and \citet{wil06} point out that galaxies with more disturbed H$\alpha$ morphologies have companions and explicitly ask whether galaxy-galaxy interactions are required to trigger the central nebulae found in cooling core BCGs. 

Evidence that recent and current galaxy interactions are an important component of BCG evolution includes the results from dark matter simulations of \citet{del07} which show that half of the BCG assembly has taken place since z$\sim$ 0.5. Observational evidence for this is also beginning to be compiled: Galaxy kinematics from  \citet{bro11} have recently confirmed that nearly equal mass pairs of two different local BCGs are gravitationally bound and likely to merge within 0.3$\,$Gyr. More generally, in a sample of 515 BCGs, \citet{liu09} find that dry major mergers are an important component of the buildup from z=0.12 to the present, but that not all the mass of the secondary is incorporated into the merger. Much of recent growth in massive galaxies may be in lower mass-ratio mergers. For example, \citet{sto11} find that BCGs have increased in size by only 30\% between z=0.25 and 1, and although \citet{ber09} and \citet{asc11} find a significant increase in BCG size between z=0.25 and 0, the BCG luminosity profile has not changed, suggesting gentle processes like minor mergers. In the theoretical work of \citet{naa09} minor mergers are also found to be an important component of the size evolution of massive galaxies, and recent observations of distant galaxies by \citet{blu11} and more local systems \citet{tru11} support these findings.

In this paper, we focus on close companions to BCGs in massive galaxy clusters. The large mass and high luminosity of BCGs allow for large luminosity ratios between the parent galaxy and companion. We use the deep photometry of the Canada-France-Hawaii Telescope Legacy Survey (CFHTLS) program to study the galaxies, so that we can probe BCGs with 20:1 luminosity ratio companions; potentially major and minor mergers. Specifically, we measure the number (N$_{c}$) and luminosity (L$_{c}$) of the companions and their proximity. Using a comparison sample of luminous galaxies gathered from the Millennium simulations, we estimate the level of contamination and calculate the dynamical friction time-scale for close companions near BCGs. The paper is organized as follows: in Section~\ref{dat} we discuss our datasets and method for calculating N$_{c}$ and L$_{c}$, in Section~\ref{res} we present our results and how they may be affected by interloping galaxies, in Section~4 we discuss the relative importance of minor and major mergers on BCG and non-BCG evolution, and in Section~\ref{con} we present our conclusions. Throughout the paper we adopt a cosmology with $\Omega$$_{m}$=0.3, $\Omega$$_{\Lambda}$=0.7 and $H_{o}$=71$\,$km$\,$s$^{-1}$$\,$Mpc$^{-1}$.

\section{Data and method}\label{dat}
\subsection{CFHTLS BCGs and controls}

\begin{figure*}
 \subfigure{
\begin{minipage}[c]{0.22\textwidth}
        \centering
        \includegraphics[width=1.5in,angle=0]{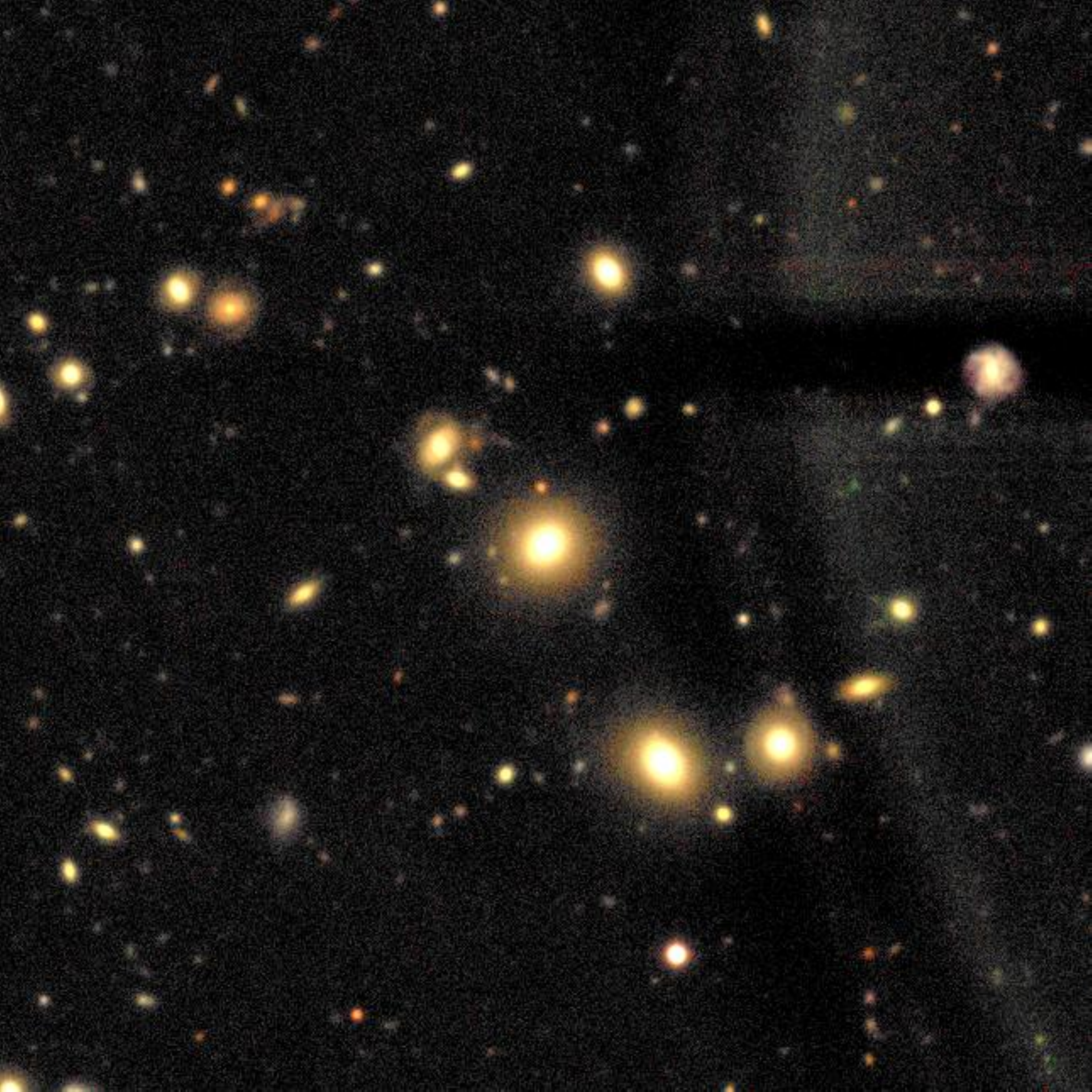}
        \end{minipage}%
    }
    \subfigure{
     \begin{minipage}[c]{0.22\textwidth}
        \centering
        \includegraphics[width=1.5in,angle=0]{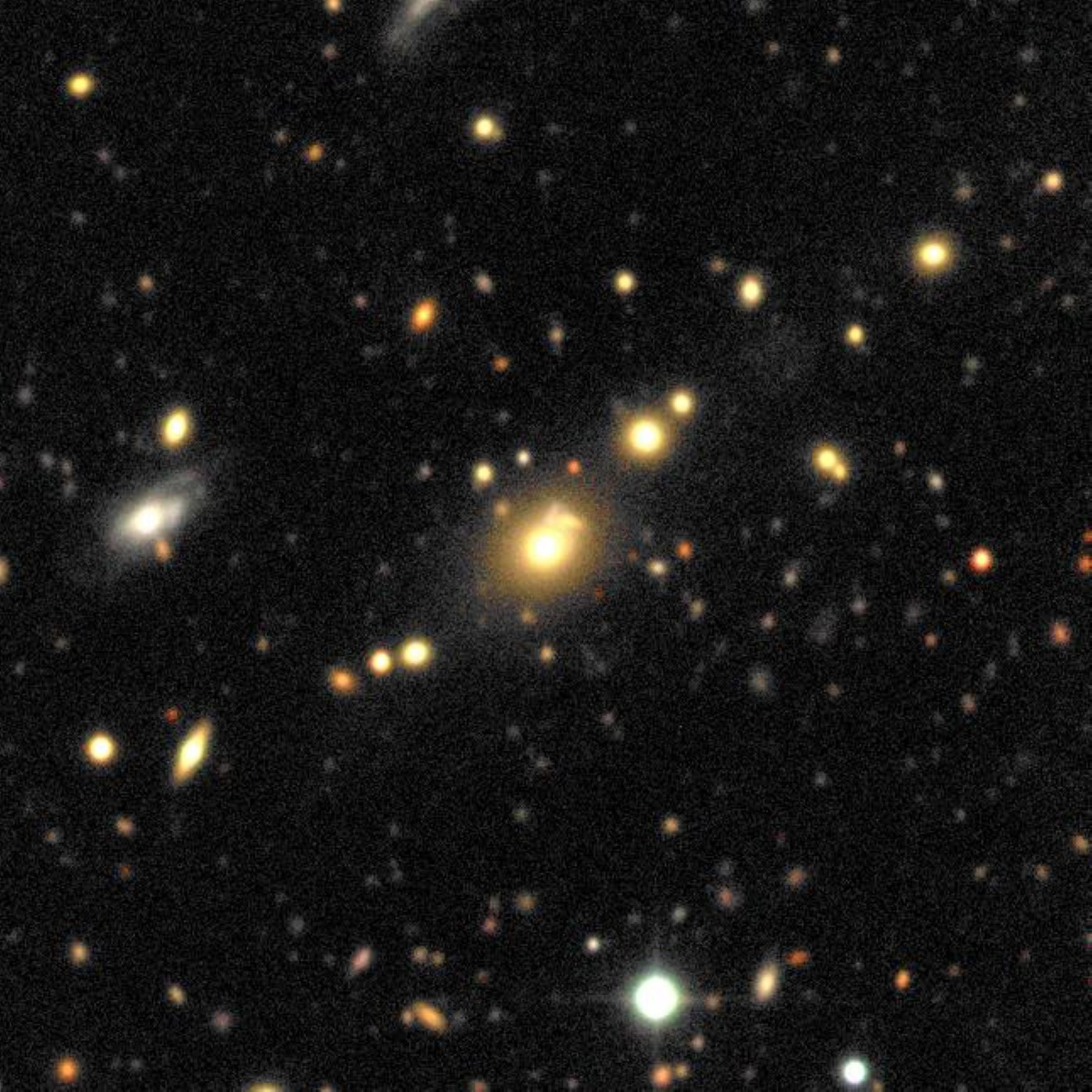}
         \end{minipage}%
    } 
    \subfigure{
     \begin{minipage}[c]{0.22\textwidth}
        \centering
        \includegraphics[width=1.5in,angle=0]{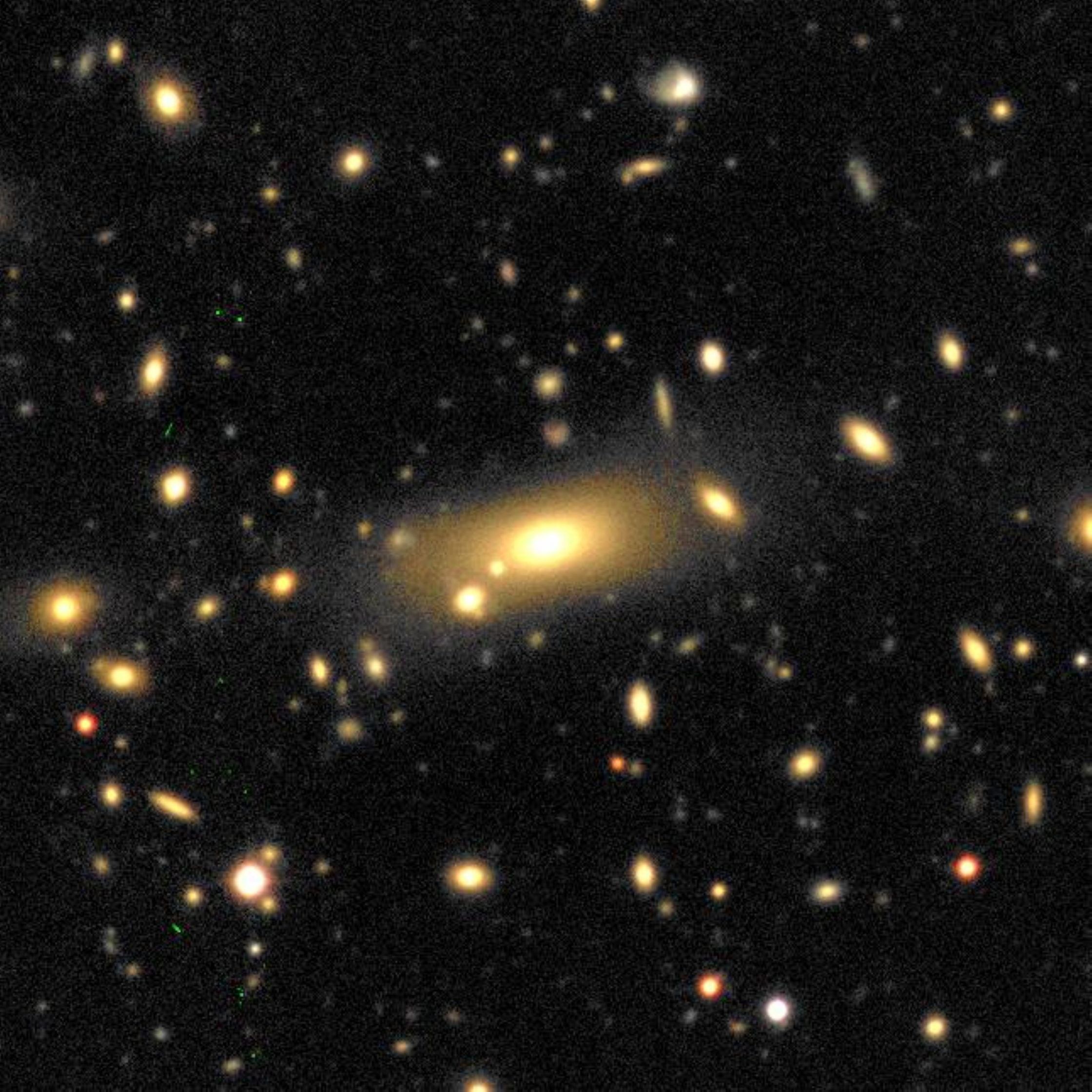}
         \end{minipage}%
    } 
    \subfigure{
     \begin{minipage}[c]{0.22\textwidth}
        \centering
        \includegraphics[width=1.5in,angle=0]{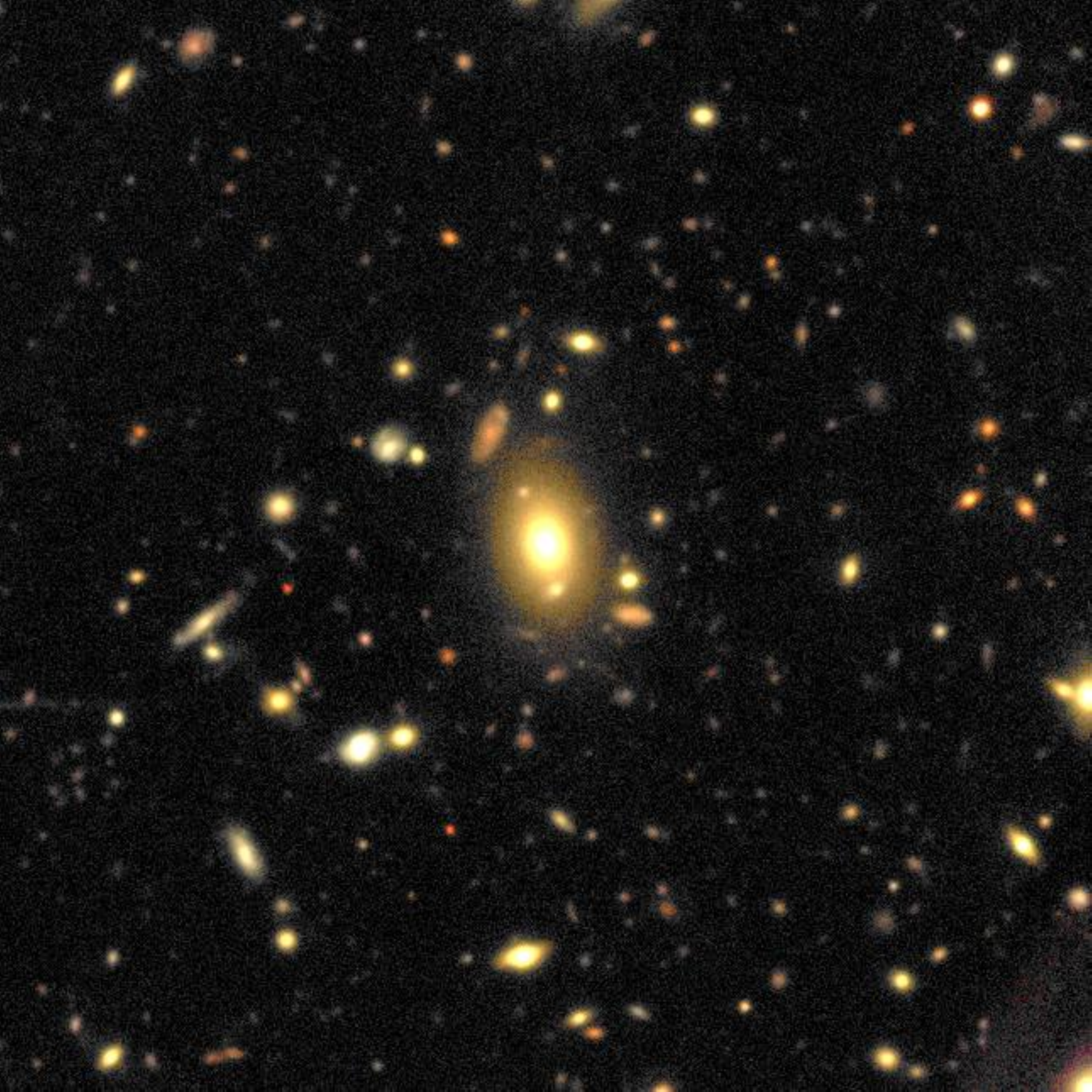}
         \end{minipage}%
    } 
    

 \subfigure{
\begin{minipage}[c]{0.22\textwidth}
       \centering
        \includegraphics[width=1.5in,angle=0]{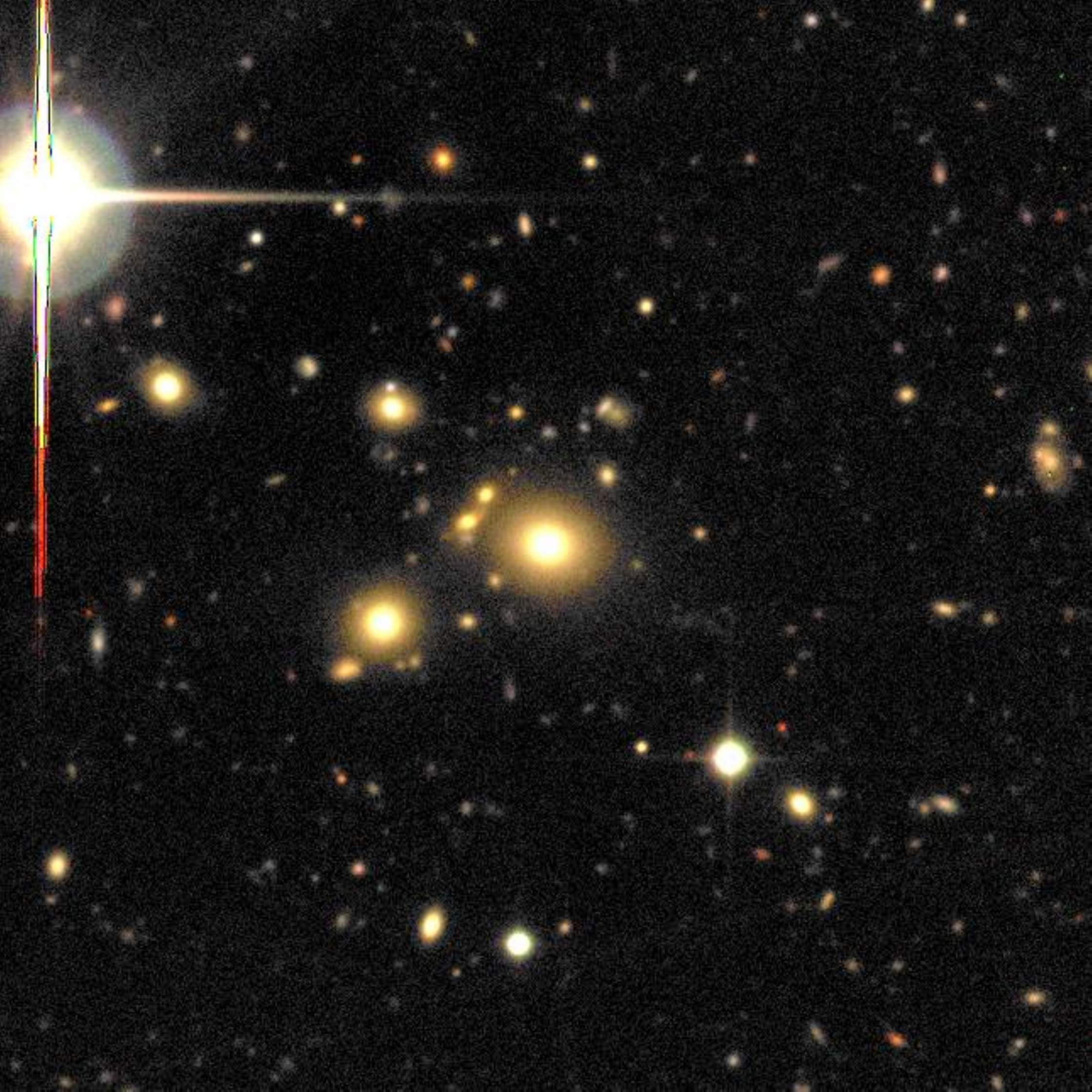}
        \end{minipage}%
    }
    \subfigure{
     \begin{minipage}[c]{0.22\textwidth}
        \centering
        \includegraphics[width=1.5in,angle=0]{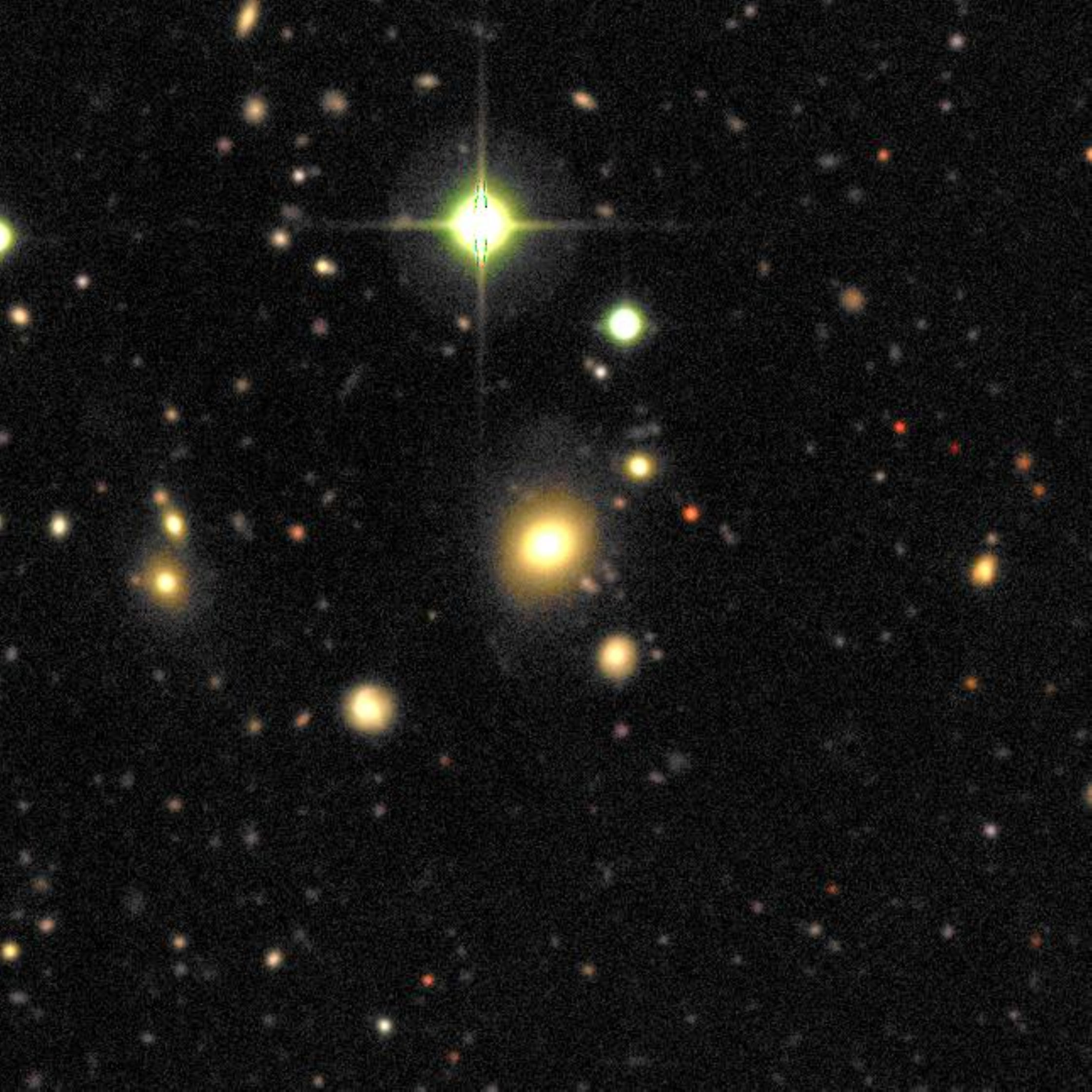}
         \end{minipage}%
    } 
    \subfigure{
     \begin{minipage}[c]{0.22\textwidth}
        \centering
        \includegraphics[width=1.5in,angle=0]{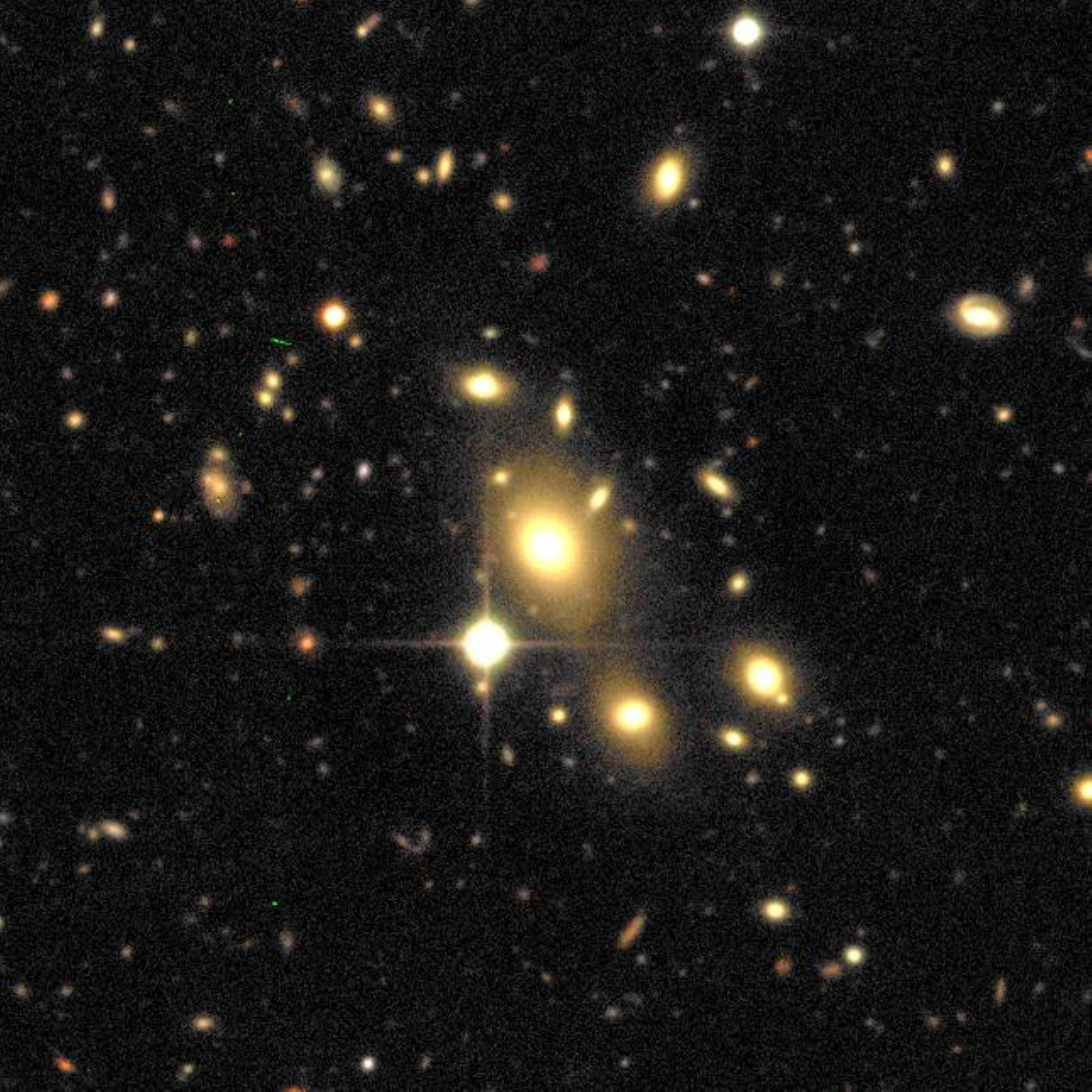}
         \end{minipage}%
    } 
    \subfigure{
     \begin{minipage}[c]{0.22\textwidth}
        \centering
        \includegraphics[width=1.5in,angle=0]{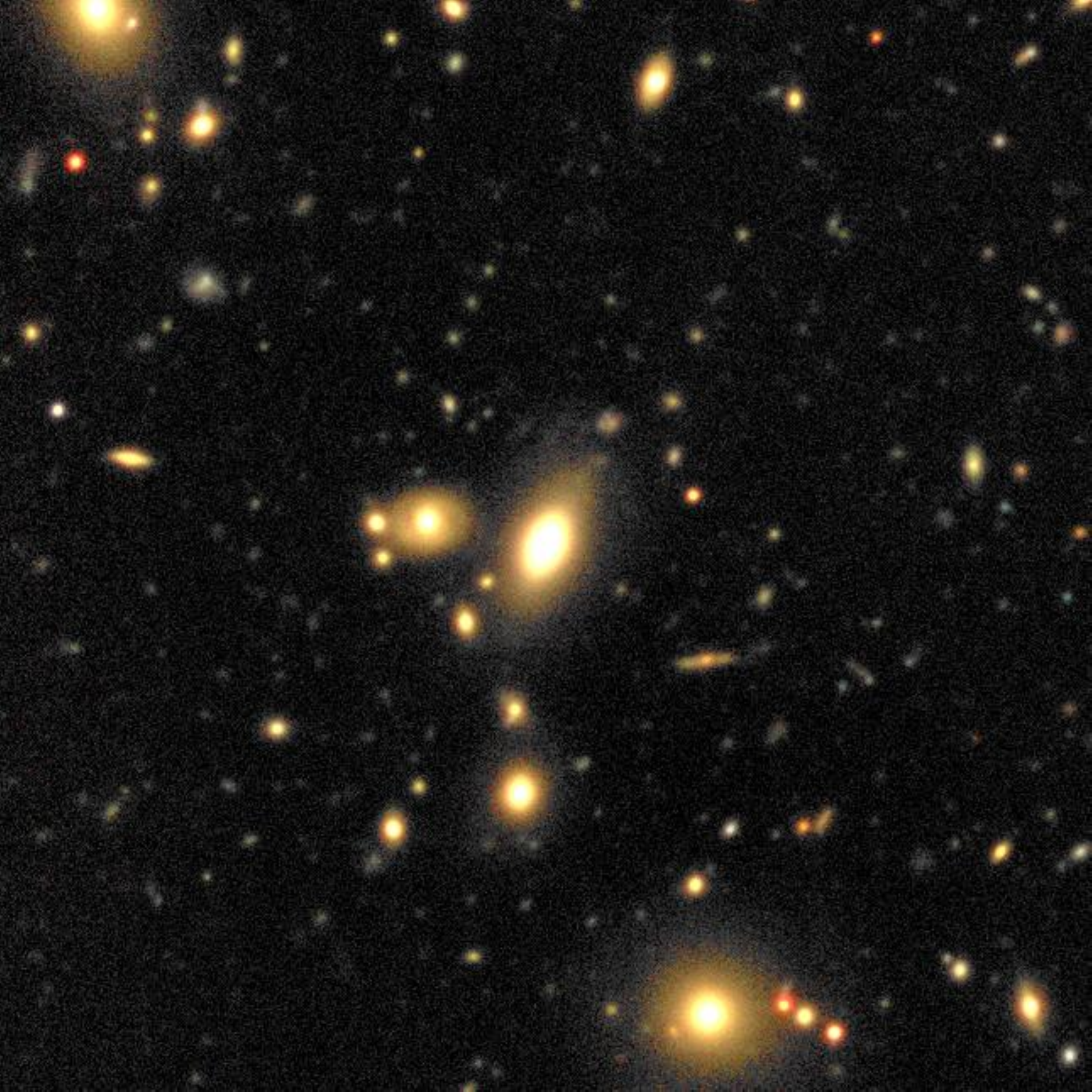}
         \end{minipage}%
    } 
\caption[\bf BCG and control colour images.] {\it Top - For illustrative purposes, we show five-colour images of four representative BCGs in our sample of 91. They display relaxed, regular morphologies, typical of BCGs. From left to right, the BCGs have (M$_{r}$,z) of (-22.86,0.25), (-22.96,0.25), (-23.45,0.24), (-23.59,0.28).  \label{colbcg} Bottom - Four galaxies from the sample of 197 luminosity-matched controls. From left to right, the BCGs have (M$_{r}$,z) of (-22.85,0.23), (-23.07,0.25), (-23.34,0.23),(-23.78,0.28). All figures are 120$^{\prime\prime}$ on a side, which corresponds to 425$\,$kpc at z=0.25, where z is the photometric cluster redshift derived from the red sequence fit. \label{colcont}}
  \end{figure*}

The BCGs and control galaxies are chosen from a list of clusters found in the CFHTLS wide fields (W1, W2, W3 and W4) data release version T0004. This release is complete for g$^{\prime}$, r$^{\prime}$, and i$^{\prime}$ in target depth, though it covers only 119 of 170 square degrees on the sky. The survey depths are 25.5, 24.8, and 24.5, in AB magnitudes for the 3 colour bands, respectively. We use the clusters from \citet{lu09} found using the red sequence method of \citet{gla05}. All redshifts in this study are the photometric redshifts derived from their analysis. There are 127 clusters in their study for which there are at least 12 red members brighter than M$^{*}$+2 magnitudes, within 0.5$\,$Mpc of the cluster centre, and which have at least 50 galaxies within 1$\,$Mpc of the cluster centre. Figures~7 and 9 of \citet{lu09} show stacked colour magnitude diagrams from the cluster sample, as well as the colour magnitude diagram for the richest cluster in the sample, respectively.

We search for potential cluster members within 1$\,$Mpc of each cluster centre. We use the publicly available photometric catalogues (from TERAPIX, and available online through the Canadian Astronomical Data Centre) and gather all available optical bands from the full T0006 data release, the first complete version of the catalogues including u$^{*}$ and z$^{\prime}$ photometry (with respective AB depth of 25.3 and 23.6). Because the clusters cover a range of redshifts between z$=$0.15-0.39, we apply the K-correction code of \citet{bla07}. 

We pick the BCG as the brightest galaxy in r$^{\prime}$ which is red ((u$^{*}$-r$^{\prime}$)$>$2.22) and also within 0.25$\,$Mpc of the cluster centre, there are 101 of these. Because this is essentially based off of the red-sequence catalog of  \citet{lu09}, we are only looking at red BCGs. Recently \citet{pip11} has found that blue BCGs make up a small fraction of the BCG population for clusters at redshifts between 0.1 and 0.3; our sample does not contain these blue BCGs.  

We construct a volume limited survey of BCGs with absolute magnitudes brighter than M$_{r}$=-20.06 (r$^{\prime}$=19.2 at z$=$0.15). By nature, BCGs are at the bright end of the galaxy luminosity function, with typical magnitudes of M$_{r}$$=$-23, thus our magnitude cuts well sample the BCG population. As the faintest BCG in the sample is actually brighter than our cut by 1.54 magnitudes, our sample essentially includes companions down to magnitude differences of 4.8. However, we choose to restrict our analysis to companions with magnitude differences of up to 3.25 (20:1 companions) since the purity of the sample degrades when looking at fainter companions, as we will see in Section~\ref{projef}.

\begin{figure*}
\centering
\epsfxsize=7in
 \epsfbox{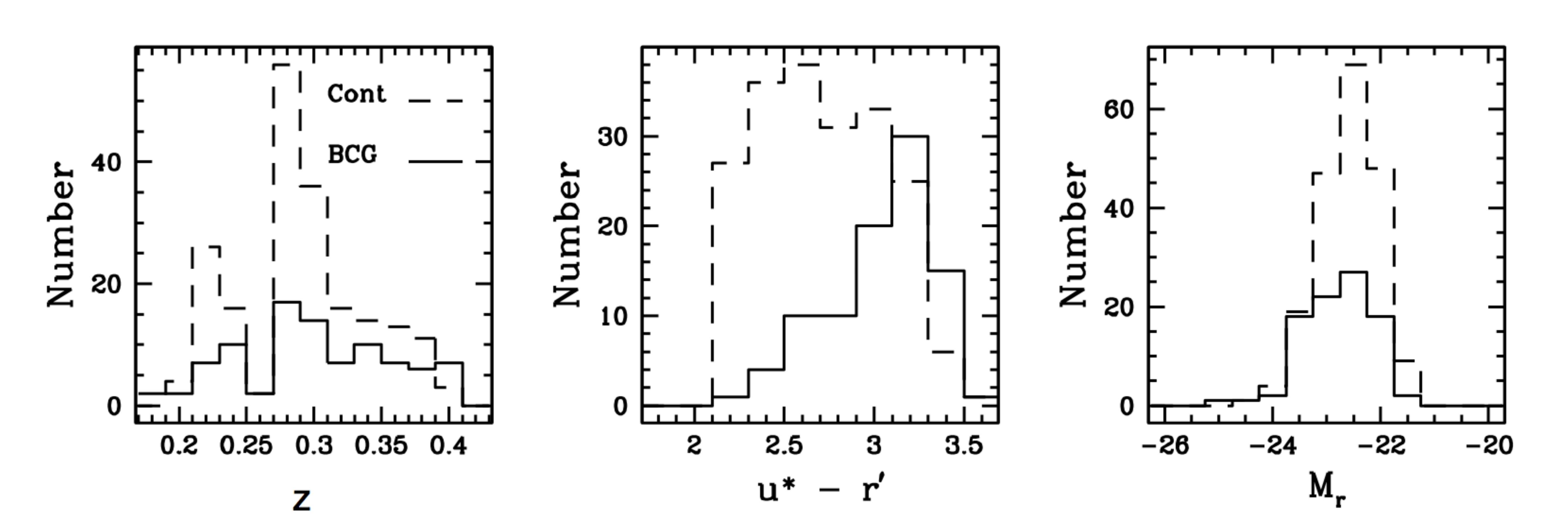} 
 \caption[bluered]{
\bf Parent galaxy properties: \it The distribution of galaxy properties for the sample of 91 BCGs (solid line) and 197 controls (dashed line) are shown. Left - The distribution of redshifts; the full redshift range has been sampled. centre - The distribution of parent galaxy colour. The BCGs are chosen to be the reddest and most luminous of the cluster, and the sample is skewed towards redder colours. The control sample is more evenly distributed. Right - The distribution of absolute M$_{r}$ magnitudes assuming the cluster redshift. By definition these are massive galaxies. Both distributions peak at M$_{r}$=-22.5. 
  \label{blueredbcg}\label{bluered}}
\end{figure*}

The sample of BCGs was examined by eye using the CFHTLS cutout service \citep{gwy08}. We found that many of the bright sources were in fact stars that had not been flagged. We re-flagged these in our galaxy tables and re-ran the BCG finder, there was one case where a star was too close to the potential BCG to accurately measure its magnitude, so we deleted this galaxy from our sample. It can happen that an obvious foreground elliptical can satisfy these criteria, and in our by-eye inspection we found 9 such cases. The final sample contains 91 BCGs. The five-colour postage stamps of four of the BCGs taken from the cutout service are shown in Figure~\ref{colbcg}. Figure~\ref{blueredbcg} shows the distribution of cluster redshifts, (u$^{*}$-r$^{\prime}$) colour, and M$_{r}$ for the BCG sample. The BCG population has a distribution of redshifts which peaks at z$\sim$0.3,  (u$^{*}$-r$^{\prime}$)$\sim$3.2, and bright magnitudes of M$_{r}$$\sim$-22.5.

We construct a luminosity matched control sample which we use to compare to the cluster BCGs. We accomplish this by collecting up to 3 other luminous red galaxies (LRGs) near the cluster centre, excluding the BCG, and including only galaxies with luminosities similar to the BCG. We also include only red galaxies on the same red sequence as the host BCG by using colour cuts of (g$^{\prime}$-r$^{\prime}$)$\pm$ 0.075 and (r$^{\prime}$-i$^{\prime}$)$\pm$0.2 \citep{lu09}. Again, all controls are inspected by eye and any lingering bright stars are removed. Figure~\ref{colcont} shows four of our 197 controls. Figure~\ref{bluered} shows the redshift, colour, and absolute magnitude distributions of the control sample. It is well matched to the BCG sample in redshift and absolute magnitude. The colours of the controls are not as heavily weighted towards the red as in the BCGs because the latter were chosen to be the reddest. 

\subsection{Identifying close companions}\label{ident}

Companions to each BCG and control galaxy are recovered by searching all galaxies in the photometric tables (red and blue) within 30$\,$kpc and 50$\,$kpc of the BCG, using the redshift derived by \citet{lu09}. The strongest evidence for interactions  and mergers in galaxy pairs has been found within 30$\,$kpc and this smaller search radius results in systems which are more likely to merge \citep{pat11,won11}. However, the larger search radius naturally leads to a larger sample of companions. Additionally, because the BCGs are the most massive systems, the larger radius of 50$\,$kpc may be more appropriate. 

We calculate the galaxy luminosity assuming the cluster redshift derived from the red sequence fitting.  We have verified that none of the close companions to the BCGs is a control, and likewise, none of the control galaxy companions is a BCG. The combination of selecting the BCG sample, which is intrinsically bright, with the deep photometry obtained with the CFHTLS therefore allows us to explore properties of close companions, ratios of 2:1, 4:1, 10:1, and down to 20:1, larger magnitude ratios than previously explored. These galaxies are those which will potentially lead to both major and minor mergers.  In Section~\ref{projef} we show the relationship between the companions we identify based on the projected separation to the parent galaxy, versus those which are close in three dimensions. Section~\ref{msims}, which follows, introduces the simulation data we use to find the 3D separations. 

\subsection{BCGs in the Millennium Simulations}\label{msims}

We use output from the Millennium Simulations, where 3D separations can be measured, in order to quantify the purity of the CFHTLS results. We use the \citet{bla05} mock redshift catalogue to do this. This catalogue contains over 5.7 million galaxies, each with apparent magnitudes, calculated stellar and virial masses, redshifts, and positions. To find a sample of simulated galaxies which mimics luminous red galaxies, we choose the $\sim$300000 with M$_{r}$$<$-24. However, most of these luminous galaxies are not at the gravitational potential of a massive cluster. Therefore, to better match to the observed data from the CFHTLS, of BCGs and controls which are embedded within massive clusters, we construct a second sample of potential CFHTLS BCG-like galaxies, and then collect the companions in the same way as for the observed data.  There are 291384 simulated red (u$^{\prime}$-r$^{\prime}$$>$2.22) and luminous (-24.5$<$M$_{r}$$<$-21.8) galaxies within the redshift range 0.2-0.4.  From this list, we keep only those that are the brightest within a radius of 0.5$\,$Mpc, resulting in 272042 galaxies. These bright galaxies have a total of 138395 companions, within 0.5$\,$Mpc. 

As alluded to by the previous numbers, the vast majority of these galaxies, especially those with the lower intrinsic luminosities, are likely to be  luminous galaxies in poor groups, or at the edge of clusters. To select BCGs that most accurately reflect those of the CFHTLS sample, we count the number of luminous red galaxies within 2 magnitudes of the potential BCG and within 0.5$\,$Mpc. Only if there are at least 3 other luminous red galaxies with velocities within 1500$\,$km/s of the BCG, we deem the potential BCG to be part of a massive cluster, 260 of these CFHTLS BCG-like galaxies remain. To find the companions to these BCGs, we search the full \citet{bla05} catalog for all galaxies, red and blue, within 0.5$\,$Mpc of the BCG. There are 871 companions.

\section{Results} \label{res}

To measure the importance of companions for BCGs and the controls,  we follow the methods of \citet{pat00}. For companions within a projected distance of 30$\,$kpc and 50$\,$kpc of the host galaxy, we count the number of  galaxies, or add up their luminosities, and divide by the total number of parent galaxies, yielding the average number (N$_{c}$) and luminosity (L$_{c}$) of companions.

There are 126 BCG companions with luminosity ratios down to 20:1 and within 50$\,$kpc (46 within 30$\,$kpc,) and 172 control companions (77 within 30$\,$kpc).  All of the companions have been inspected by eye, and obvious stars removed. The average N$_{c}$ for BCGs is 1.38$\pm$0.14,  larger than for the luminosity-matched control sample, which has N$_{c}$=0.87$\pm$0.08. The average L$_{c}$ of BCGs, 2.14$\pm$0.31$\times 10^{10}$L$_{\odot}$, is also higher than that for the non-BCG sample, 1.48$\pm$0.20$\times 10^{10}$L$_{\odot}$, where the uncertainties are calculated from the standard error on the mean. At face value, these results indicate that BCGs accrete more companions and more stellar mass more often that non-BCGs. We will return to this  issue after addressing contamination by interlopers (non-merging companions) in Section~\ref{projef}.

\subsection{The number and luminosity of companions as a function of luminosity ratio}

\begin{figure}
 \subfigure{
 \begin{minipage}[c]{0.2\textwidth}
        \centering
       \includegraphics[width=2.0in,angle=0]{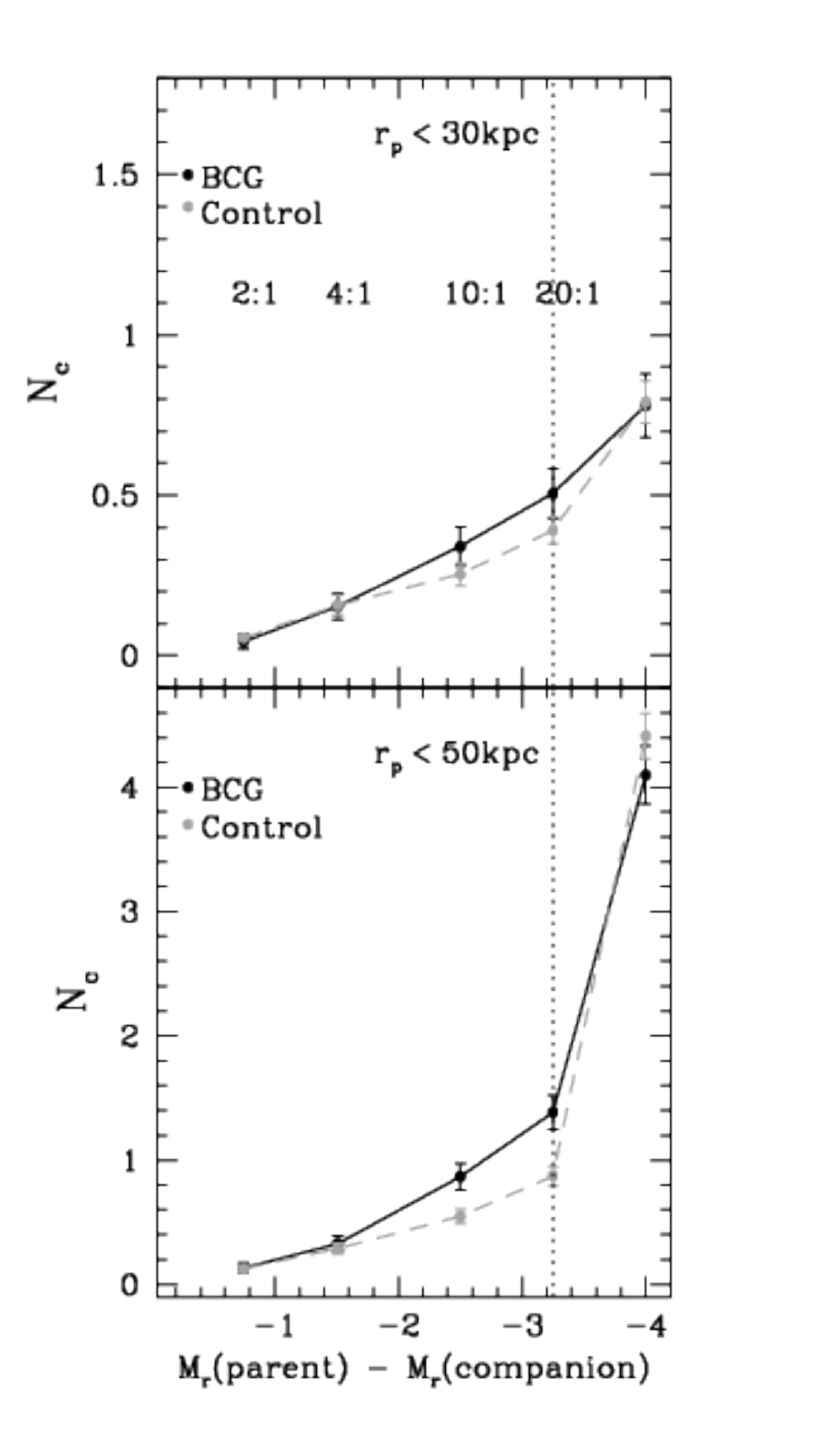}
       \end{minipage}%
    }
    \subfigure{
     \begin{minipage}[c]{0.35\textwidth}
        \centering
        \includegraphics[width=2.0in,angle=0]{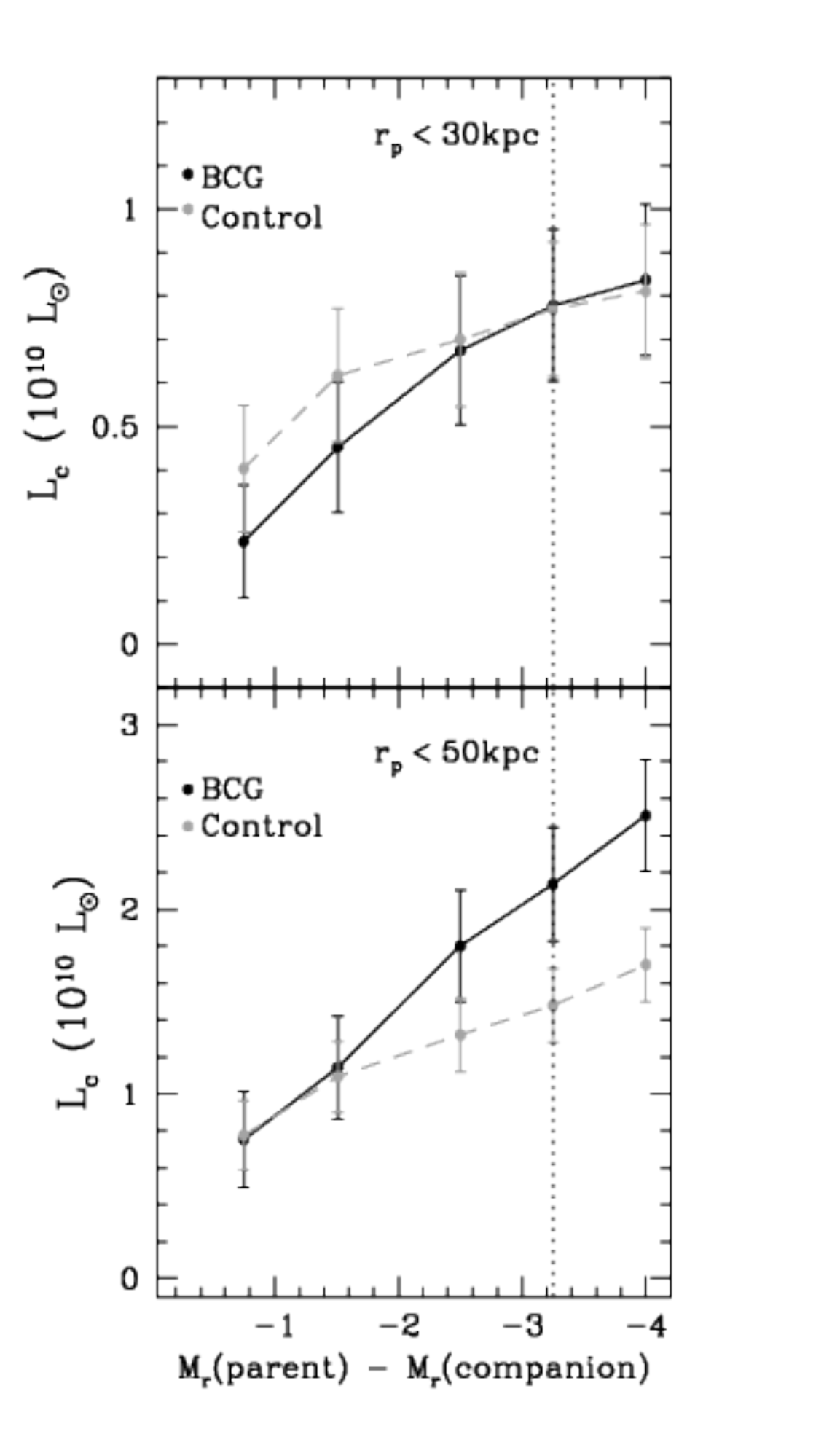}
         \end{minipage}%
    }  
\caption[\bf N$_{c}$ and L$_{c}$] {\it Left - N$_{c}$ as a function of the luminosity ratio for companions to BCGs (black) and controls (grey).  \label{MRNc} Right - L$_{c}$ as a function of the luminosity ratio for companions to BCGs (black) and controls (grey). Companions within 30$\,$kpc  are shown on the top. Those within 50 $\,$kpc are shown on the bottom where BCGs have higher values of N$_{c}$ and L$_{c}$ than the control sample. The last point includes the impure set of all measured companions beyond magnitude ratios of 20:1. \label{MRLc}}
  \end{figure}

We construct the cumulative distribution of N$_{c}$ and L$_{c}$ as a function of the luminosity ratio between the parent (BCG or control) and the companion.

Figure~\ref{MRNc} shows that the BCGs have similar low numbers of 2:1 and 4:1 companions as do the controls.  The dearth of near mass companions is expected as these luminous cluster galaxies are at the bright end of the cluster luminosity function and rare, thus less likely to have a close companion with a similar size. The trends for BCGs and controls are the same to within errors for the sample of companions within 30$\,$kpc. It is interesting, however, that the number of companions for BCGs rises more steeply than for the controls, and is consistently higher. In the sample that is statistically more significant, including companions out to 50$\,$kpc, the number of 10:1 and 20:1 companions is larger for the BCGs than for the controls.

This steeper rise for companions of BCGs can also be easily seen for L$_{c}$ (Figure~\ref{MRNc}, right panels). There is more luminosity in companions per BCG than in companions per control when integrating over all companions found, including those with ratios beyond 20:1.

\subsection{The number and luminosity of companions as a function of parent galaxy absolute magnitude}

\begin{figure}
 \subfigure{
 \begin{minipage}[l]{0.2\textwidth}
        \centering
       \includegraphics[width=2.0in,angle=0]{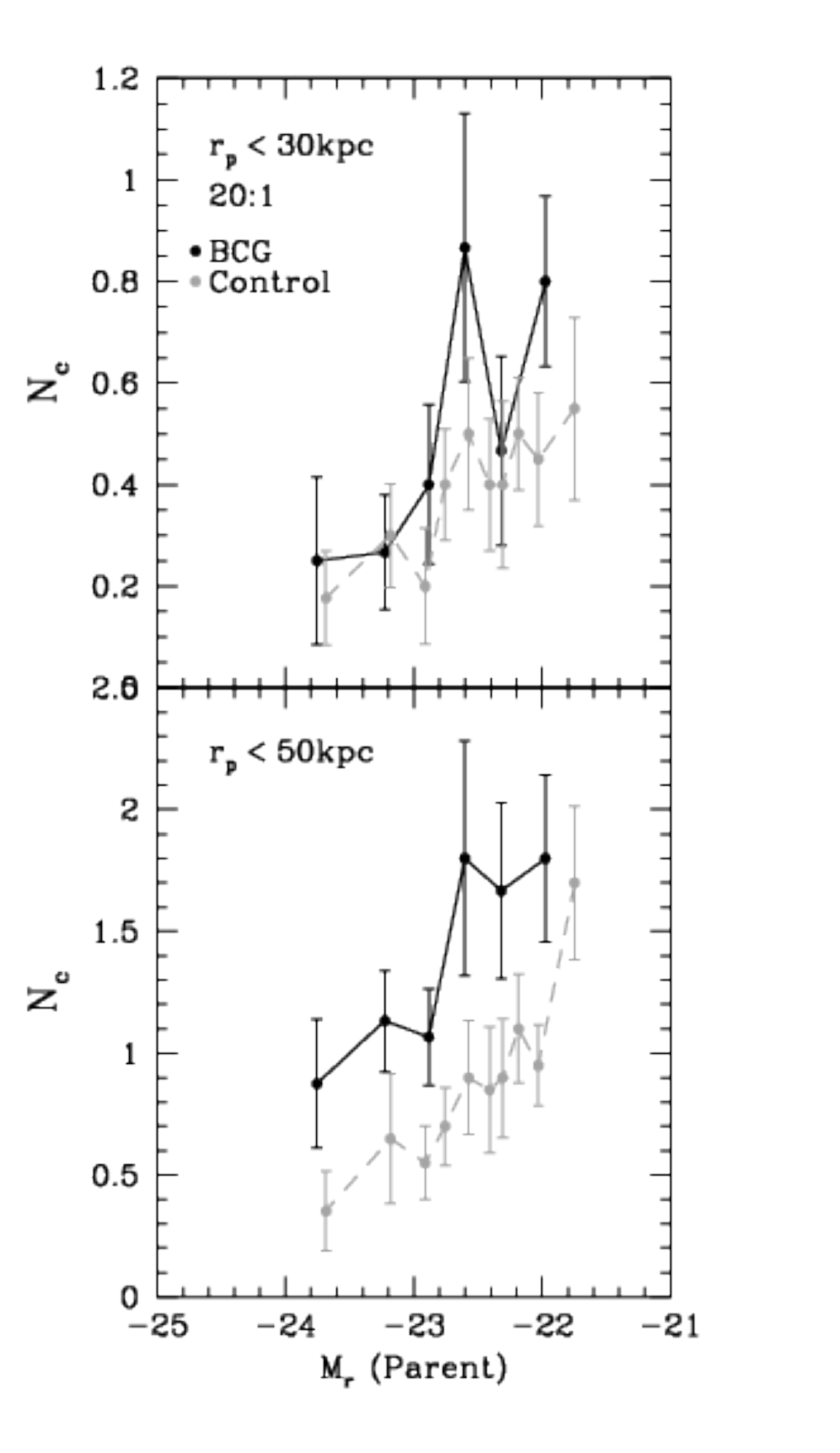}
       \end{minipage}%
    }
    \subfigure{
     \begin{minipage}[l]{0.35\textwidth}
        \centering
        \includegraphics[width=2.0in,angle=0]{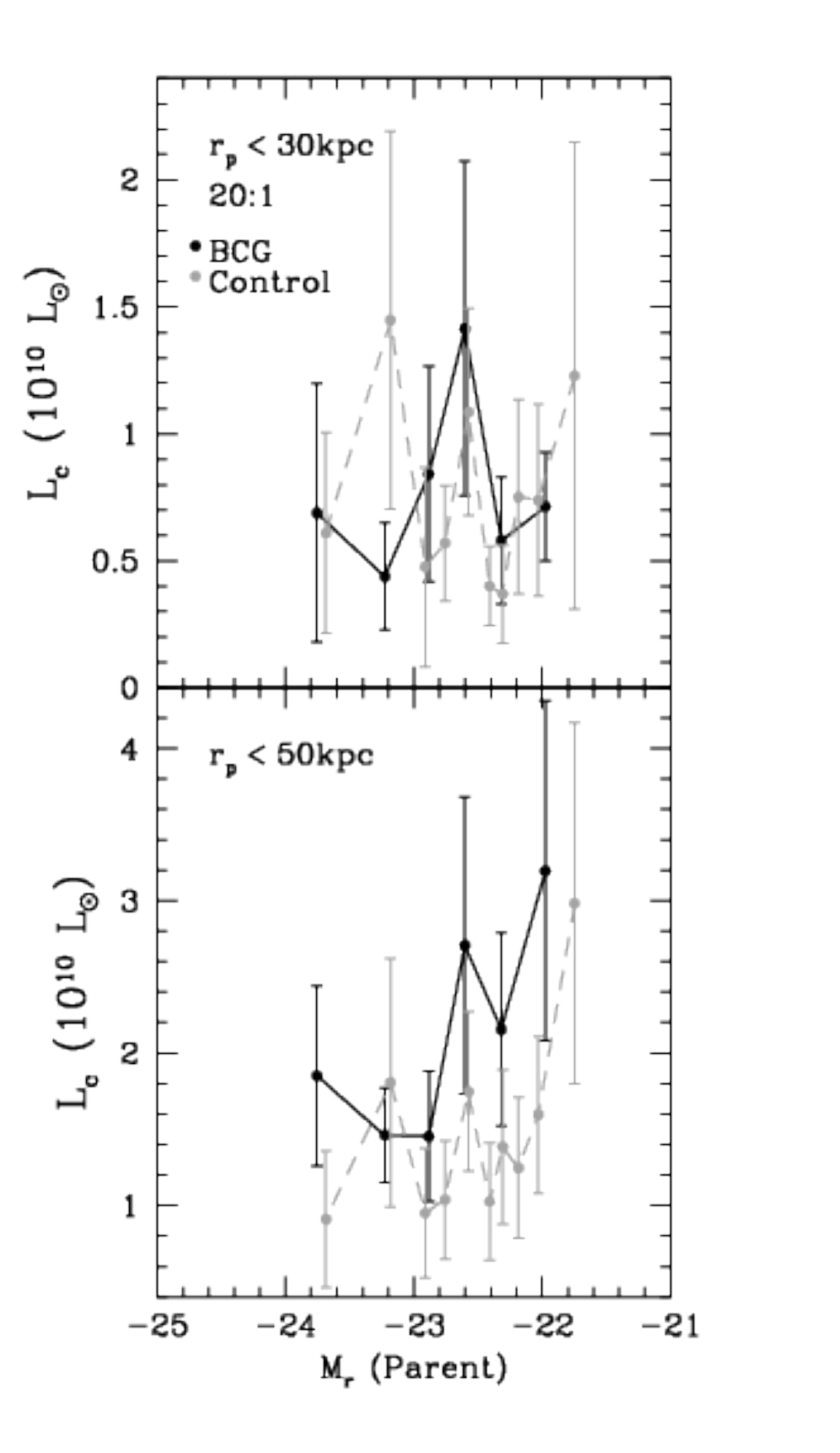}
         \end{minipage}%
    }  
\caption[\bf N$_{c}$ and L$_{c}$] {\it Left - N$_{c}$ as a function of the magnitude of BCGs (black) and controls (grey).  \label{MpNc} Right - L$_{c}$ as a function of the magnitude of BCGs (black) and controls (grey). The sample with r$_{p}$$<$30$\,$kpc is shown on the top, and the sample with r$_{p}$$<$50$\,$kpc is shown on the bottom. All panels include companions down to luminosity ratios of 20:1. \label{MBLc} }
  \end{figure}

The number of companions per BCG is plotted as a function of the absolute r$^{\prime}$ magnitude of the parent galaxy in the left panel of Figure~\ref{MBLc}. This includes all companions down to 20:1 luminosity ratios.  Clearly, this is higher for BCGs than for luminosity matched controls. Also, there is a tendency for the lower luminosity galaxies (both BCGs and controls) to have more 20:1 companions than the brighter galaxies. This last, however, may simply be because we probe fainter into the luminosity function for the fainter host galaxies. 

\subsection{The number and luminosity of companions as a function of companion absolute magnitude}

\begin{figure}
 \subfigure{
 \begin{minipage}[l]{0.2\textwidth}
        \centering
       \includegraphics[width=2.0in,angle=0]{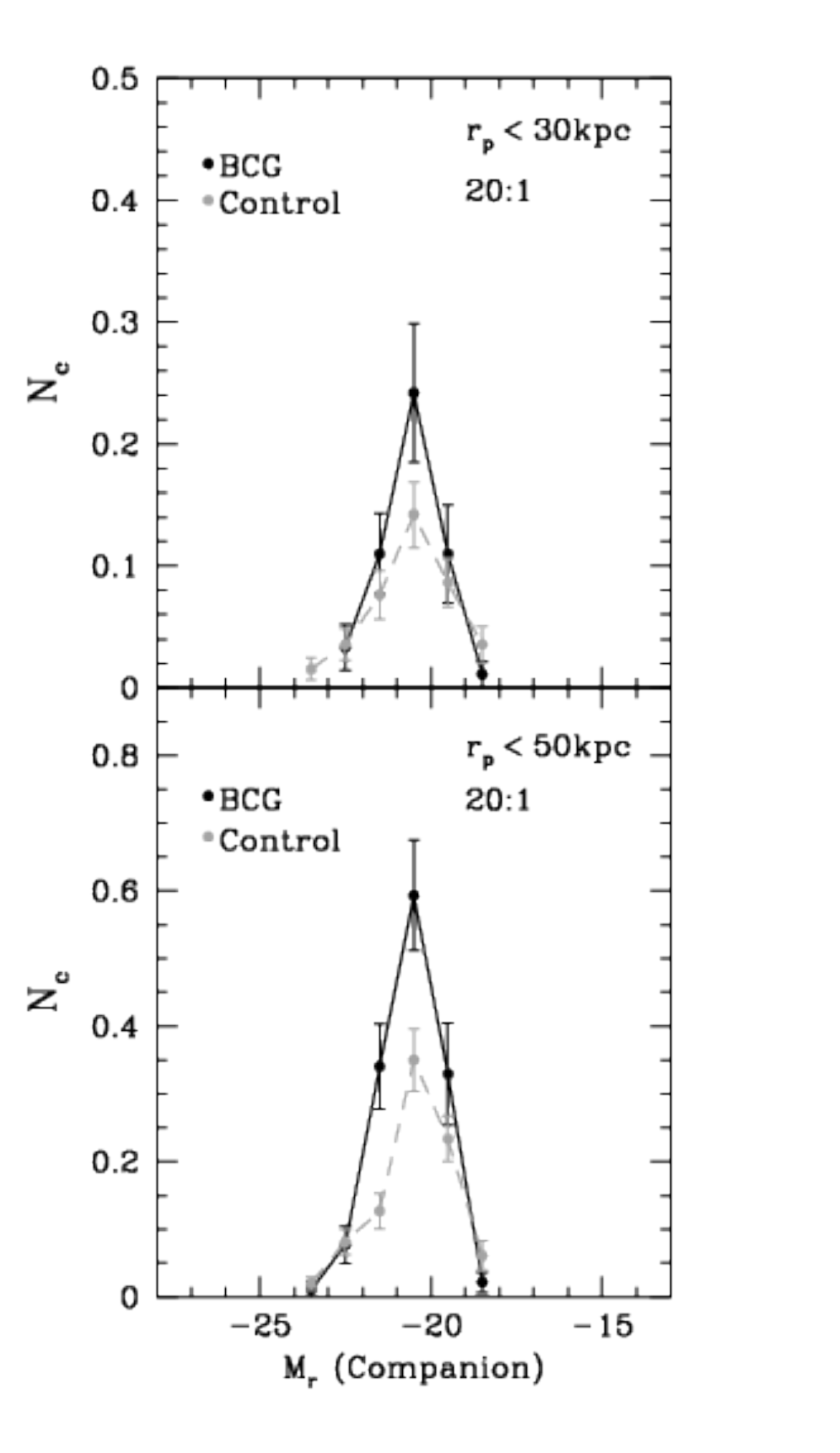}
       \end{minipage}%
    }
    \subfigure{
     \begin{minipage}[l]{0.35\textwidth}
        \centering
        \includegraphics[width=2.0in,angle=0]{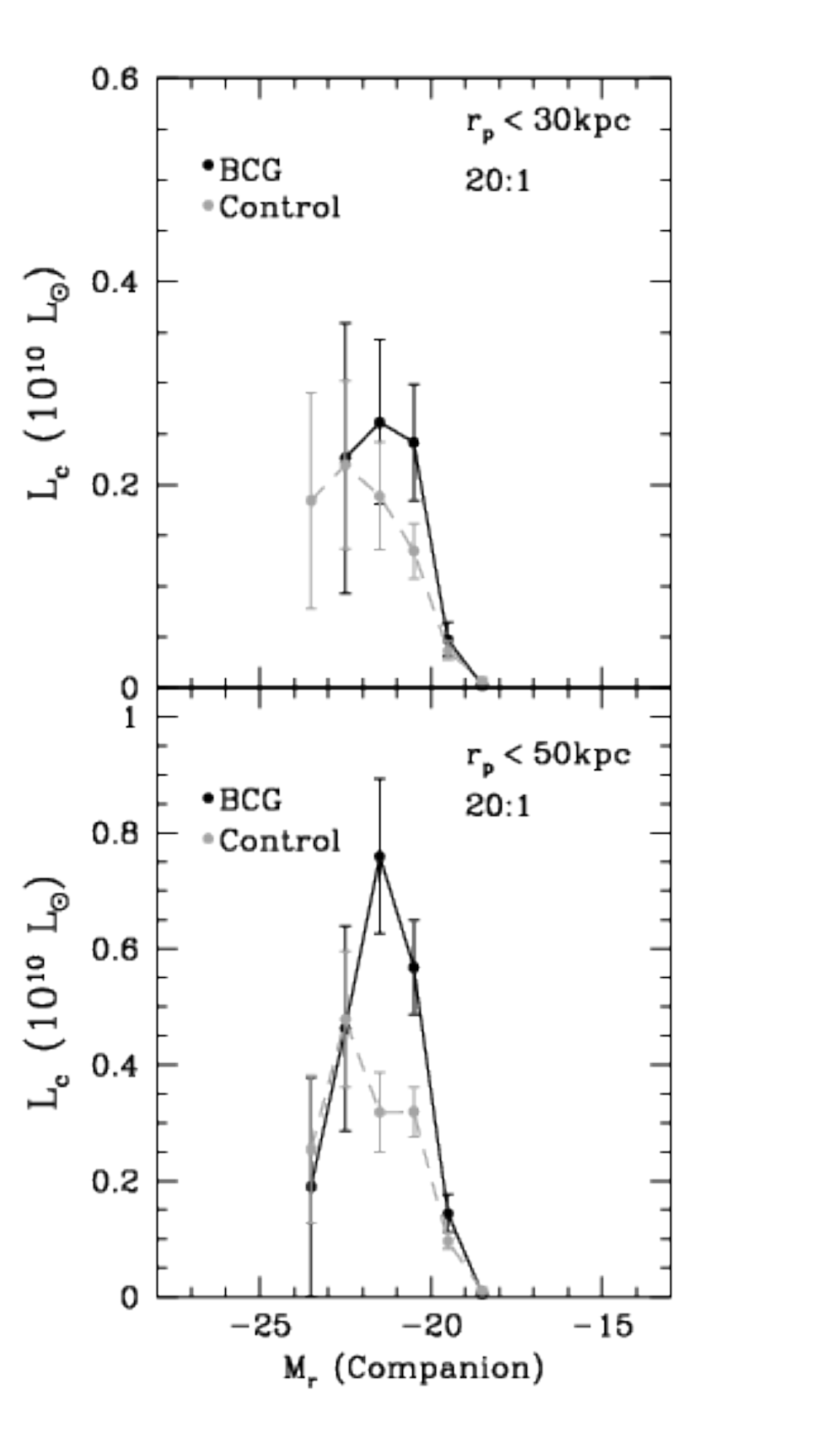}
         \end{minipage}%
    }  
\caption[\bf N$_{c}$ and L$_{c}$] {\it Left - N$_{c}$ as a function of the companion magnitude for BCGs (black) and controls (grey).  \label{MpNc} Right - L$_{c}$ as a function of the companion magnitude for BCGs (black) and controls (grey). The sample with r$_{p}$$<$30$\,$kpc is shown on the top, and the sample with r$_{p}$$<$50$\,$kpc is shown on the bottom. All panels include companions down to luminosity ratios of 20:1. \label{MpLc} }
  \end{figure}

On the left panels of Figure~\ref{MpLc}, we present the average number of companions per parent galaxy (either BCG or control) in bins of companion absolute r$^{\prime}$ magnitude including all companions down to luminosity ratios of 20:1. The histograms include all companions down to luminosity ratios of 20:1 across clusters of all redshifts. The figures clearly show that there are few very bright as well as very faint companions, with the peak of the distribution of companion magnitudes at M$_{r}=$-20.5. 

The luminosity matched control sample peaks at the same companion magnitude as the BCG sample. Although, within 50$\,$kpc the number of control companions with \mbox{M$_{r}$$\sim$$\,$-21} is fewer than for the BCGs. N$_{c}$ for companions out to only 30$\,$kpc is the same to within errors for BCGs and controls; a Kolmogorov-Smirnov test shows no evidence that the BCG and control companions were drawn from different distributions. 

As with N$_{c}$, the peak in companion luminosity for BCGs is higher than for the controls when including companions out to 50$\,$kpc. Again, out to only 30$\,$kpc, the two curves are the same within errorbars, and a Kolmogorov-Smirnov test shows no evidence of a different parent population. These results are consistent with the overall N$_{c}$ values reported at the beginning of this section.

Overall, most companion galaxies have moderate magnitudes of $\sim$$\,$-21. This distribution is most likely a function of our parent galaxy sample selection, which peaks at  \mbox{M$_{r}\sim$$\,$-23}, and no BCGs with magnitudes fainter than -21.6 are in our sample thus no 20:1 companions with magnitudes fainter than -18.4 exist in our sample. Since the companions are all chosen to have magnitude ratios within 20:1, the distribution in companion mass is also expected. Fainter companion galaxies certainly exist, but will exist in the less-populated bins of the lower-magnitude parent galaxies (see Figure~\ref{blueredbcg}, right panel). To illustrate this, we calculated N$_{c}$ and L$_{c}$ for only the subset of BCGs with magnitudes between -22.75 and -23.25. The sample size in this bin is low, with only 22 BCGs, however, within the limited magnitude range, the number of companions does in fact rise as the companion mass falls.  Figure~\ref{MpLc} reminds us that the bulk of BCG companions we explore in this study do not exist as dwarf galaxies (at least not down to 20:1), nor as equal-mass pairs, but rather in normal L$_{*}$ cluster members.

\subsection{Accounting for projection effects} \label{projef}

\begin{figure*}
 \subfigure{
\begin{minipage}[c]{0.33\textwidth}
        \centering
        \includegraphics[height=2.8in,angle=0]{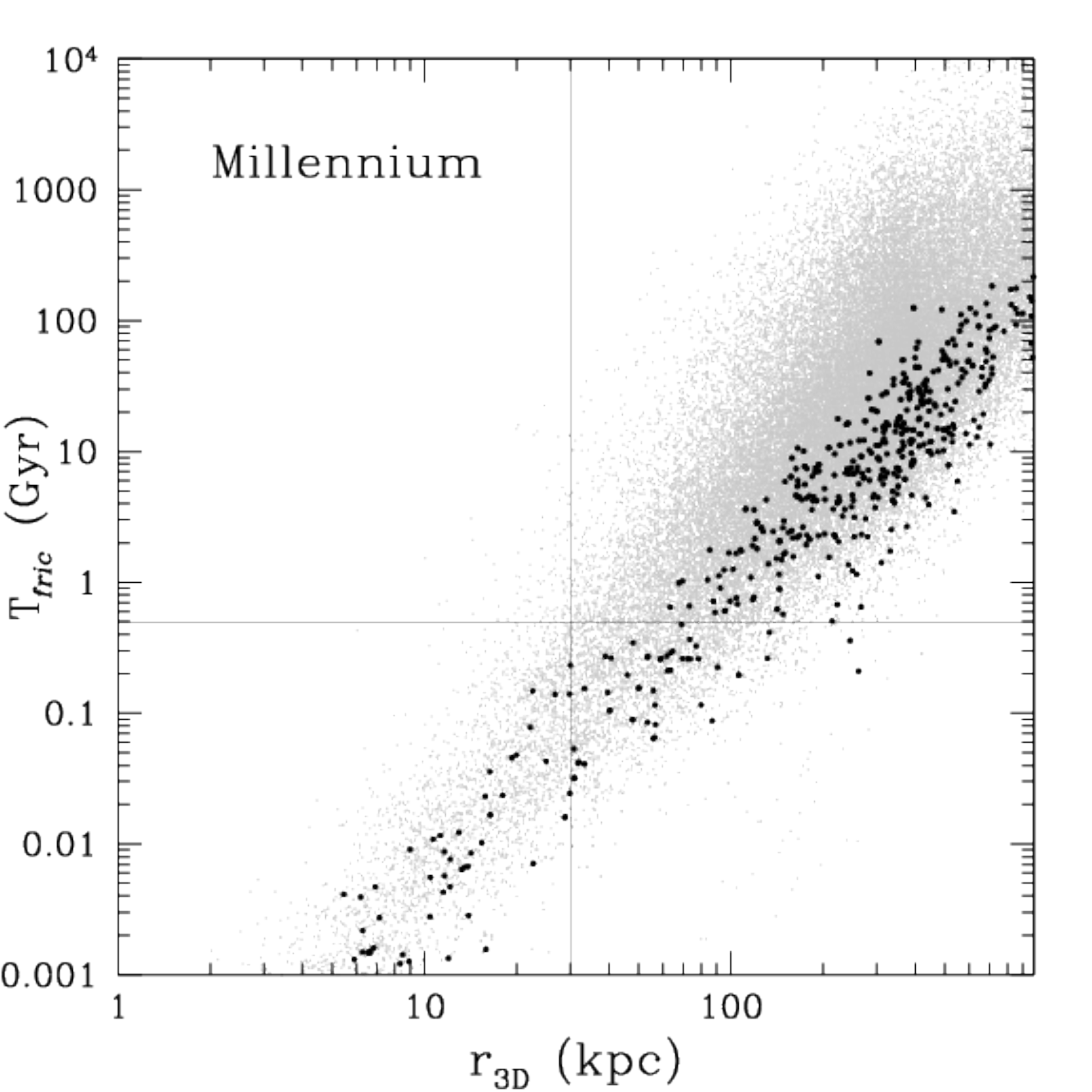}
        \end{minipage}%
    }
    \subfigure{
     \begin{minipage}[c]{0.33\textwidth}
        \centering
        \includegraphics[height=2.8in,angle=0]{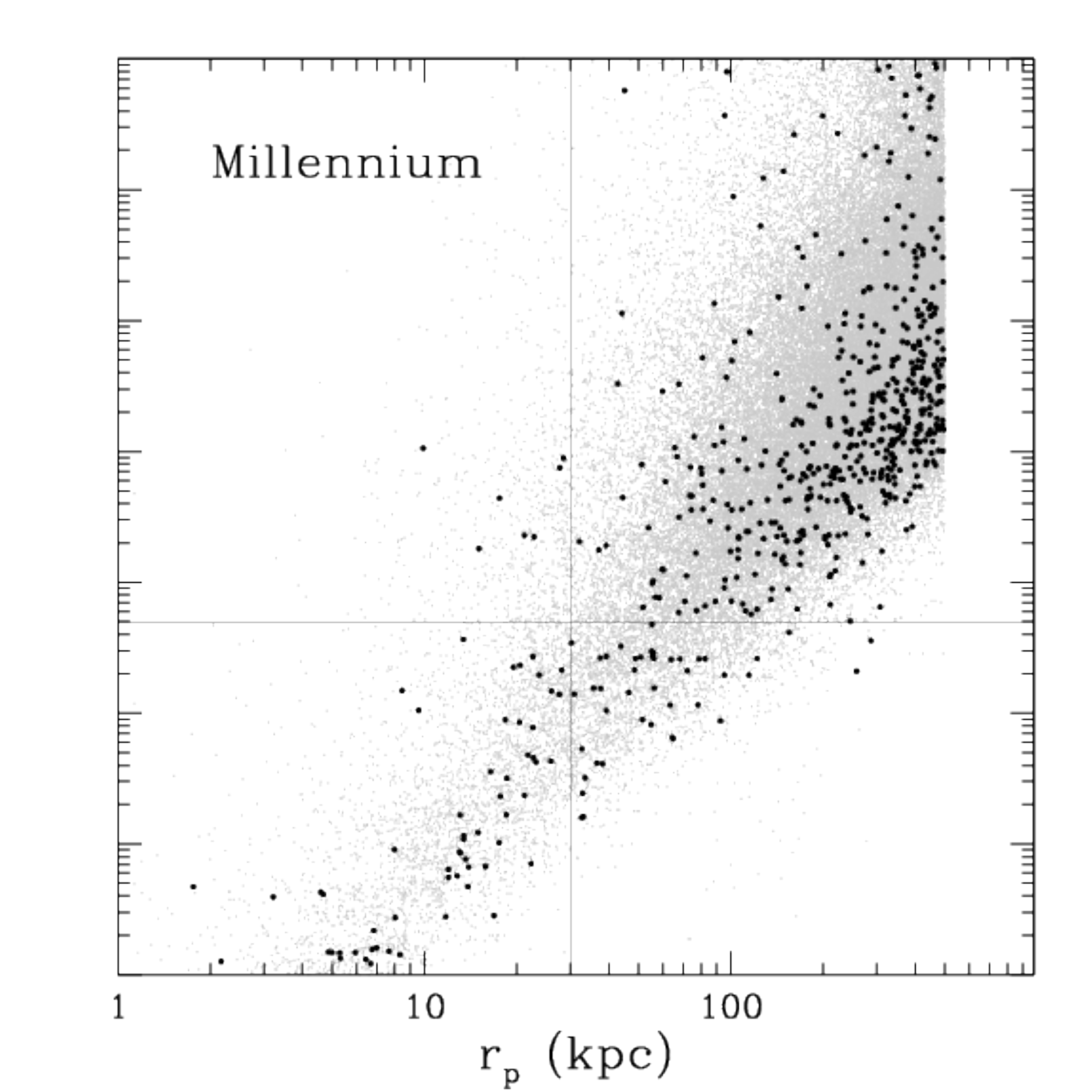}
         \end{minipage}%
    } 
        \subfigure{
     \begin{minipage}[c]{0.3\textwidth}
        \centering
        \includegraphics[height=2.8in,angle=0]{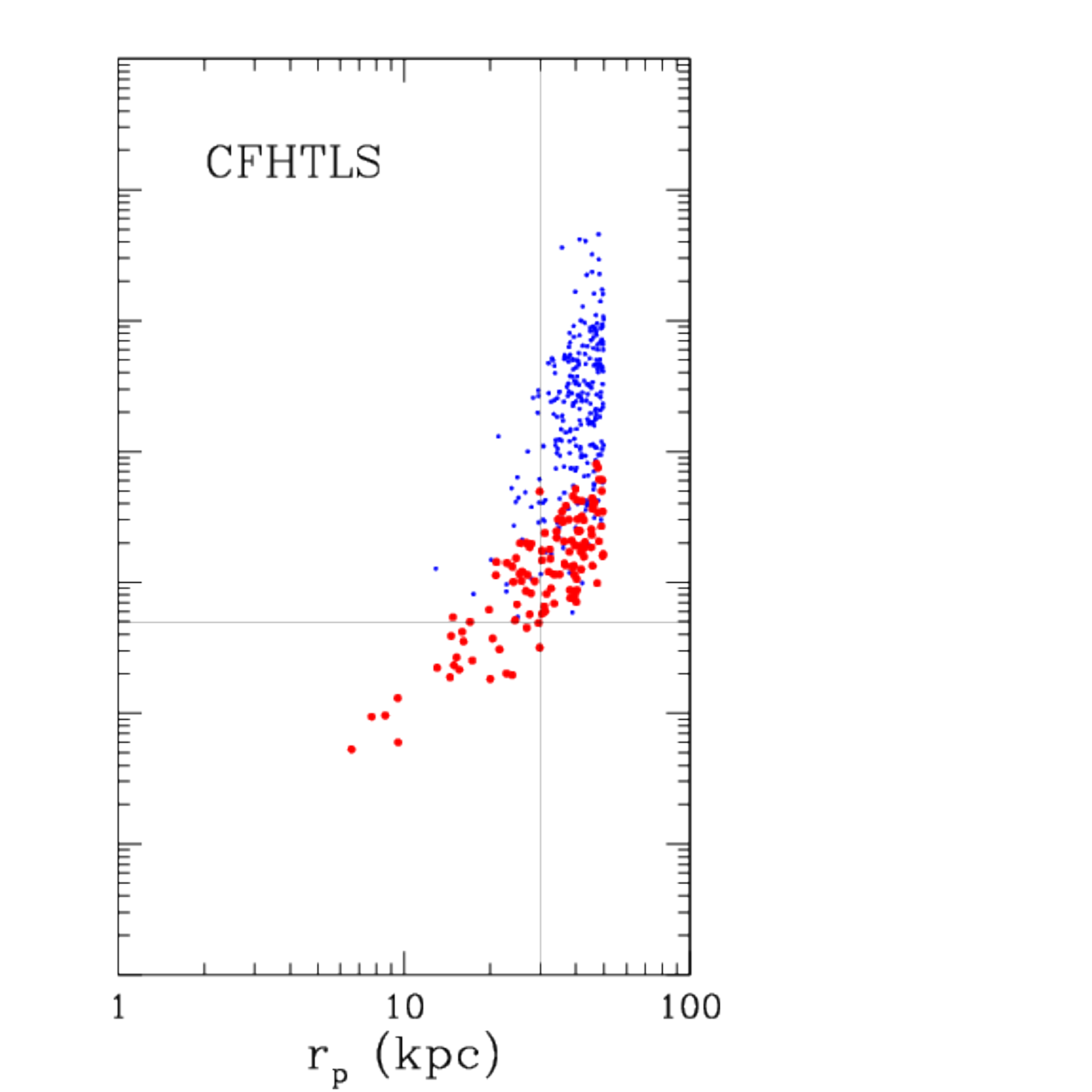}
         \end{minipage}%
    } 
    \caption[Dynamical time-scale]{{\bf The dynamical friction time-scale:} {\it The dynamical friction time-scale plotted as a function of the parent-companion separation.  The first two panels are for the Millennium simulation data. Left -  The real 3D separation and circular velocity is used to calculate T$_{fric}$. Centre - The projected separation is used to calculate T$_{fric}$. In the Millennium data, the grey points are all luminous galaxies and the black dots are the CFHTLS BCG-like galaxies described in the text. Right - T$_{fric}$ calculated for the CFHTLS data, using the projected separation and v$_{c}$ = 800$\,$km/s, typical of cluster velocity dispersions. The 126 red dots are companions down to 20:1 magnitude differences, and the smaller blue dots are the additional 250 (incomplete set of) even fainter companions.\label{milsim}} }
\end{figure*}

Throughout the analysis of the observational data, we have studied the number and luminosity of companions which are close to the parent galaxy on the sky, not the true 3D separation of the companion and parent. This leads to the possibility of contamination by foreground and background objects, erroneously observed as part of the 2D projection of the galaxies on the sky.  We can, however, turn to dark matter simulations in order to study the importance of these projection effects for small separations. We define f$_{3D}$, which is the fraction of companions which are close in 3D, thus allowing for the removal of interlopers.

Galaxies which are close in three dimensions are those likely to merge with the BCG. We attempt to separate the potential merging companions from unrelated foreground and background galaxies using the \citet{bla05} mock redshift catalogue, based on the output of the Millennium Simulations. After calculating the dynamical friction time-scale, we analyse the companions of two samples of simulated galaxies defined in Section~\ref{msims} in order to quantify the projection effects. Both parent samples match the observed data in luminosity, colour, and redshift. They are: (1) All luminous red galaxies and (2) the CFHTLS BCG-like galaxies which are further selected to mimic the cluster core environment. 

\subsubsection{The dynamical friction time-scale}

There has been much discussion in recent literature about the variety of merger time-scales possible. Many authors have found that several parameters are important when looking at the merging time-scales including gas fractions \citep{lot11} and mass ratios \citep{wil11,lot11}. Based on the same Millennium Simulation data, \citet{kit08} found that dynamical friction time-scales depend on redshift, galaxy mass, as well as projected separation. In this work, we do not consider the gas mass fraction of the potential mergers, however, we do take care to calculate dynamical friction time-scales based on a sample of simulated galaxies which has been closely matched to our observed data.


Following \citet{pat00}, we estimate the merger time-scale using the dynamical friction time-scale for our samples of high luminosity galaxies. We follow the method of \citet{bin87}, assuming circular orbits and a dark matter density profile of $\rho$(r)$\propto$ r$^{-2}$.

\begin{equation}
$$T$_{fric}$=2.64$\times$10$^{5}$ r$^{2}$v$_{c}$/M ln($\Lambda$),$$
\end{equation}

where the physical pair separation is r in kpc, v$_{c}$ is the circular velocity in km$\,$s$^{-1}$, M is the mass of the companion galaxy in solar masses, and ln $\Lambda$ is the Coulomb logarithm. We use the value of ln $\Lambda\sim$ 2, measured by \citet{dub99}, since this value agrees with simulations of the orbital decay of equal mass mergers. 

We find that even though the galaxies we study likely reside at the centre of rich clusters, where the relative velocities are often thought to be too high for merging to occur, the dynamical friction time-scale is indeed short for a large fraction of the companions. This means that close pairs are likely to be bound and lead to real mergers.

In Figure~\ref{milsim}, we show that the dynamical friction time-scale, T$_{fric}$ \citep{bin87}, is $<$ 0.5$\,$Gyr, for close companions within r$_{3D}$ $\sim$ 30$\,$kpc. Ninety percent of companions to the luminous galaxies, with projected separations $<$ 30 kpc, have T$_{fric}$ $<$ 0.5$\,$Gyr. All of the CFHTLS BCG-like simulated galaxies within r$_{3D}$ $\sim$ 50$\,$kpc have short merger time-scales (left panel). These calculations also show that 50\% of companions with short merging times are rejected from our measurements of N$_{c}$ and L$_{c}$ as they have projected separations $>\,$50$\,kpc$ (middle panel).

For comparison, we calculate T$_{fric}$ for the CFHTLS data. We do not have redshift data for the CFHTLS galaxies, therefore we use a value of v$_{c}$=800$\,$km/s, the median value of the 3D velocity difference in the Millennium CFHTLS BCG-like galaxies and their companions with luminosity ratios between 20:1 and 2:1. This value is also consistent with the velocity dispersions of massive galaxy clusters, where BCGs are located. To find the mass, we turn to the literature for characteristic BCG values. \citet{van91} determined the average M/L$_{R}$ for elliptical galaxies to be 4.7, and recent results from \citet{cap06} find typical M/L$_{R}$ of 5.8 for luminous galaxies with M$_{R}$ between -22 and -25. We take the average of these two values and use a mass to light ratio of 5.3. Using this ratio, we convert the luminosity to a mass. The trend appears similar to what we found for the Millennium data, with a short enough T$_{fric}$ for $\sim$60\% of the 20:1 companions within 30$\,$kpc to merge with their parent galaxy in under 0.5$\,$Gyr (Figure~\ref{milsim}, right panel). With the caveats that the observational data is based on only 126 BCG companions and our assumption on the velocities, we conclude that within the small distances probed here the dynamical friction time-scale is short, and most companions will likely merge with their parent.

\subsubsection{The fraction of real 3D companions}

\begin{figure}
\centering
        \includegraphics[width=3.05in,angle=0]{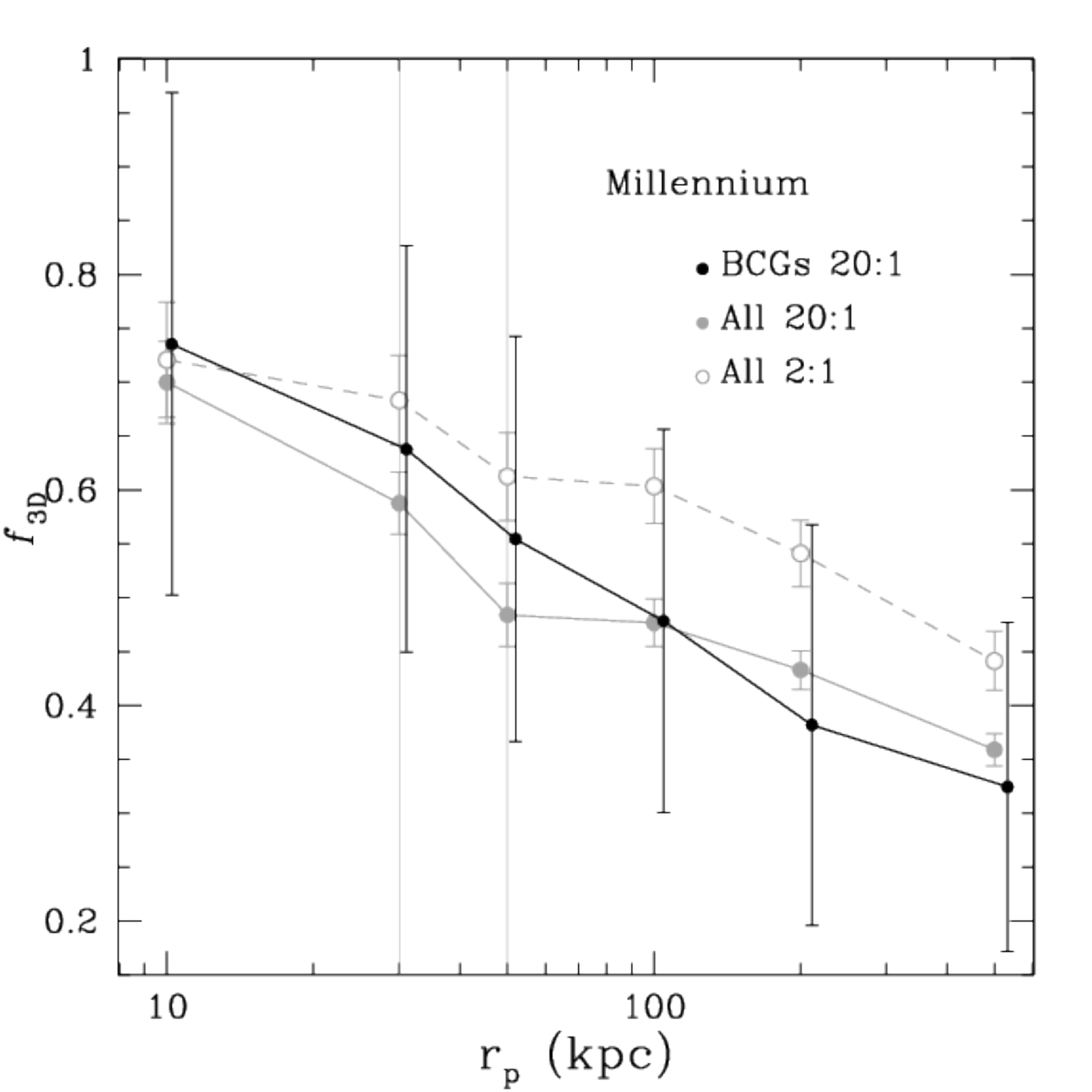}
    \caption[Interloper contamination]{ {\bf Interloper contamination:} {\it The fraction of real 3D companions, f$_{3D}$, is shown as a function of projected separation. The amount of interloper contamination is similarly small for 20:1 (filled grey dots) and 2:1 (open grey dots) companions to luminous galaxies as well as for the CFHTLS BCG-like galaxies (filled black dots, includes all 20:1 companions) at small r$_{p}$. f$_{3D}$ falls below 50\% within projected separations of 50$\,$kpc for 20:1 companions, and 300$\,$kpc for 2:1 companions (essentially BCG-LRG pairs). Vertical lines highlight distances of 30$\,$kpc and 50$\,kpc$.\label{f3d}} }
\end{figure}

Figure~\ref{f3d} shows the fraction of real 3D companions (f$_{3D}$) as a function of the projected distance. The amount of interloper contamination is similarly small for 20:1 and 2:1 companions with the closest r$_{p}$. The simulated data also show that while f$_{3D}$ is only $\sim$0.35 for 20:1 companions at r$_{p}$ $\sim$ 500$\,$kpc, f$_{3D}$ jumps to $\sim$0.50 for companions within a projected distance of 50$\,$kpc, and to $\sim$0.70 within a projected separation of 10$\,$kpc. This trend of declining f$_{3D}$ with projected radius is similar for the smaller sample of CFHTLS BCG-like galaxies. For the 2:1 companions, the amount of contamination is even less severe, with a greater than 50\% purity for projected distances lower than 300$\,$kpc. We do not show the curve for 2:1 companions to the CFHTLS BCG-like sample since the errors are too large due to the small sample size. When quoting our final N$_{c}$ in the conclusion, we apply the f$_{3D}$ from the statistically most significant 20:1 sample. 

\section{Discussion}

\subsection{BCG formation and evolution}
\subsubsection{The importance of minor mergers}
Because BCGs are so extreme in size and few in number, it is rare to find companions with luminosity ratios of 2:1 and higher. Our sample of BCGs and controls both show a similar value of N$_{c}$ as does the more inclusive sample of galaxy surveys studied by \citet{pat08}. Constraining their analysis to a luminosity ratio of 2:1, these authors find a value of N$_{c}$ $\sim$0.02 for primary galaxies of magnitudes \mbox{-22 $<$ M$_{r}$ $<$ -18}. Despite the fact that their sample includes less massive galaxies in a range of environments, with a range of colours, our study finds slightly larger results for BCGs with characteristic magnitudes of M$_{r}$ $\sim$ -22.5, but consistent to within errors  (N$_{c}$ = 0.04$\pm$0.02, including companions within 30$\,$kpc).

This analysis adds to the mounting evidence supporting the importance of high-ratio companions (potential minor mergers) for galaxy formation. We have shown that the luminosity in companions per BCG does not plateau, but continues to increase even out to mass ratios of 20:1.  There is a significant amount of luminosity that exists in the sum of lower mass companions - the luminosity in companions with luminosity ratios of 4:1 is 1.14$\pm$0.28$\times 10^{10}$L$_{\odot}$, whereas it has almost doubled to 2.14$\pm$0.31$\times 10^{10}$L$_{\odot}$ when including 20:1 companions. This suggests that minor mergers are at least as important as major mergers in building the mass of the BCG. This is particularly true for the companions out to 50$\,$kpc, with the caveat that they suffer more contamination from interlopers. The same is true for the control sample. Although the L$_{c}$ in BCG companions rises more steeply than for control companions, essentially the luminosity of the smaller companions is important for both. Moreover, \citet{blu11} have recently shown that for a sample of distant luminous galaxies (1.7$<$z$<$3) half of the mass resulting from mergers results from major (1:4) mergers, and half from minor (1:100). The building up of the galaxy mass through mergers is consistent with the results of \citet{ber09} and \citet{asc11}, who find BCGs that are more local, have larger sizes, but that the BCG luminosity profiles are not disturbed, implying a smooth accumulation of mass.

\subsubsection{Downsizing}
It is known that more luminous BCGs tend to lie in more massive clusters \citep[e.g.][]{edg91,bro08}, however, this does not seem to be dominating the number of close companions, since it is the {\it less} luminous parent galaxies that have the {\it larger} number of companions. This naturally leads to a fainter integral companion luminosity for the brighter parent systems, which is consistent with what downsizing predicts. If the light traces mass, and the brighter BCGs represent more massive systems, and if the larger mass systems formed first, it is consistent that we would observe more companions
in the smaller systems, which could still be actively building up. However, to test this hypothesis, the effect of the dwarf galaxies that our study may be missing should be considered. First, the luminosity function is not probed to as faint of magnitudes for the brighter BCGs, and second, the larger haloes of the brighter galaxies may preferentially cover adjacent faint companions. We will explore this in a following paper. 

\subsubsection{Evolution with cosmic time}
There are several recent studies regarding merging among local samples of large red galaxies, or galaxies specifically in groups  \citep[e.g.][]{mci08,mas09,liu09}. The most similar sample to the present study is that of \citet{liu09}, who examined BCGs, at redshifts z$<$0.12 with companions within projected separations of 30$\,$kpc, and luminosity ratios of 4:1. They find 49/515 have close pairs (N$_{c}$ $\sim$0.1), and once folding in morphological information (tidal tails, etc) deem 18 of these to be major mergers (N$_{c}$ $\sim$0.03). To compare our results to theirs, we look at N$_{c}$ for BCGs,  including only the most luminous companions, those within 4:1. We find a higher value of N$_{c}$=0.15$\pm$0.03 for the moderate redshift sample ($\sim$0.1 after correcting for interlopers). Although there is a larger number of companions in our higher-redshift sample, it is important to note that the methodology in both studies is very different, and also that \citet{liu09} include only the red companions. The present study includes all companions regardless of colour (though few of the low-ratio companions are blue.) However,  if the close companions do merge with the BCG and are not replaced, one would expect N$_{c}$ to decrease as the universe evolves, which is consistent with the difference in N$_{c}$ between our massive cluster galaxies, and those of \citet{liu09}. 

\section{Conclusions and Future Work} \label{con}

By measuring N$_{c}$, L$_{c}$, for a sample of observed galaxies, and calculating the dynamical friction time-scale, this work indicates that the assembly of the very luminous galaxies through mergers, both major and minor, is an important process in galaxy evolution at the moderate redshifts of z=0.15-0.39. This is true for both the BCGs, as well as a sample of luminosity matched non-BCG cluster galaxies. However, we do find that  probing down to luminosity ratios of 20:1, BCGs have more companion galaxies, and more luminosity in their companions than do the non-BCGs. 

From examining massive galaxies in the Millennium Simulations, we find that projection effects within the small separations we are dealing with are low. For example, for BCG companions within 50$\,$kpc and magnitude ratios down to 20:1, the correction due to projection effects is $\sim$50\%, so that N$_{c}$$\sim$0.7 and L$_{c}$$\sim$1.0$\times 10^{10}$L$_{\odot}$. For companions within 30$\,$kpc,  the correction factor is $\sim$60\% and thus, N$_{c}$$\sim$0.3 and L$_{c}$$\sim$0.9$\times 10^{10}$L$_{\odot}$.

Finally, we find that the dynamical friction time-scale at these distances is short enough for merging to take place. This leads us to the conclusion that the luminosity in bound merger candidates down to luminosity ratios of 20:1 could be adding as much as $\sim$10\% to the stellar mass of a 5$\times$10$^{11}$M$_{\odot}$ BCG, over only 0.5$\,$Gyr. This is likely to be an underestimate since there are bound to be some galaxies beyond 50$\,$kpc that will merge with the BCG. These results are true not only for BCGs, but also for the control sample of luminous red cluster galaxies. We conclude that the addition of the mass of the close companions to the total mass of the parent galaxy is an important component of evolution in bright cluster galaxies. Including the bound companions out to 50$\,$kpc could help resolve the contradiction between recent measurements which show no luminosity evolution in BCGs since z=1.4  \citep{whiley08,col09,sto10}, and the semi-analytic models of \citet{del07}, who find BCGs to have recently undergone significant hierarchical growth.

  \section*{Acknowledgments}

We thank T. Lu for early access to her cluster catalogue. DRP gratefully acknowledges the receipt of an NSERC Discovery Grant, which helped to fund this research. We would also like to thank C. Lopez-Sanjuan and the anonymous referee each for their useful comments which have lead to a significantly improved version. This study is based on observations obtained with MegaPrime/MegaCam, a joint project of CFHT and CEA/DAPNIA, at the Canada-France-Hawaii Telescope (CFHT) which is operated by the National Research Council (NRC) of Canada, the Institut National des Science de l'Univers of the Centre National de la Recherche Scientifique (CNRS) of France, and the University of Hawaii. This work is also based in part on data products produced at the Canadian Astronomy Data Centre as part of the Canada-France-Hawaii Telescope Legacy Survey, a collaborative project of NRC and CNRS. The Millennium Simulation data bases used in this paper and the web application providing online access to them were constructed as part of the activities of the German Astrophysical Virtual Observatory.

\bibliography{le1}

 \end{document}